\documentclass[sigconf, nonacm]{acmart}

\usepackage{cleveref}
\usepackage{multirow}
\usepackage{verbatim}
\usepackage{xfrac}
\usepackage{graphicx}
\usepackage{dcolumn}
\usepackage{bm}
\usepackage{listings}
\usepackage{tabularx}
\usepackage{booktabs}
\usepackage{lipsum}
\usepackage{todonotes}
\usepackage{tikz}
\usepackage{url}
\usepackage{hyperref}

\usepackage[outdir=./figures/pdf/]{epstopdf} 

\usepackage{subcaption}
\lstdefinestyle{custompython}{
  belowcaptionskip=1\baselineskip,
  breaklines=true,
  frame=L,
  xleftmargin=\parindent,
  language=Python,
  showstringspaces=false,
  basicstyle=\footnotesize\ttfamily,
}

\newcommand{\cclass} [1] {\textnormal{\textsf{#1}}}

\definecolor{babyblue}{rgb}{0.54, 0.81, 0.94}
\definecolor{forestgreen(web)}{rgb}{0.13, 0.55, 0.13}

\newcolumntype{Y}{>{\centering\arraybackslash}X}

\usepackage{dcolumn,booktabs}
\newcolumntype{d}[1]{D{.}{.}{#1}}

\graphicspath{{figures/}}

\usepackage[ruled, vlined, linesnumbered, nofillcomment]{algorithm2e}

\DontPrintSemicolon
\SetKwInOut{Input}{Input}
\SetKwInOut{Output}{Output}

\DeclareMathOperator{\argmin}{argmin}
\DeclareMathOperator{\argmax}{argmax}

\newcommand{\optproblemdef}[3]{
	\begin{center}
	\begin{minipage}{0.95\columnwidth}
		\noindent
		#1
		\vspace{10pt}\\
		\setlength{\tabcolsep}{3pt}
		\begin{tabularx}{\textwidth}{@{}lX@{}}
			\textbf{Input:}     & #2 \\
			\textbf{Task:}  & #3
		\end{tabularx}
	\end{minipage}
	\end{center}
}

\title{Fast algorithms to improve fair information access in networks}

\author[]{Dennis Robert Windham$^1$,\hspace{1em} Caroline J. Wendt$^1$,\hspace{1em} Alex Crane$^2$,\hspace{1em} Madelyn J Warr$^2$,\hspace{1em} Freda Shi$^2$,\hspace{1em} Sorelle A. Friedler$^3$,\hspace{1em} Blair D. Sullivan$^2$,\hspace{1em} Aaron Clauset$^{1,4,5}$\hspace{1em}}

\affiliation{
\institution{\vspace{1em}
$^1$Department of Computer Science, University of Colorado\\
$^2$University of Utah\\
$^3$Haverford College\\
$^4$BioFrontiers Institute, University of Colorado\\
$^5$Santa Fe Institute}
\country{}}

\begin{document}
\pagestyle{plain}

\begin{abstract}
We consider the problem of selecting $k$ seed nodes in a network to maximize the minimum probability of activation under an independent cascade beginning at these seeds.
The motivation is to promote fairness by ensuring that even the least advantaged members of the network have good access to information.
Our problem can be viewed as a variant of the classic influence maximization objective, but it appears somewhat more difficult to solve: only heuristics are known.
Moreover, the scalability of these methods is sharply constrained by the need to
repeatedly estimate access probabilities.\looseness=-1

We design and evaluate a suite of~$10$ new scalable algorithms which crucially do not require probability estimation. To facilitate comparison with the state-of-the-art, we make three more contributions which may be of broader interest.
We introduce a principled method of selecting a pairwise information transmission parameter
used in experimental evaluations, as well as a new performance metric which allows for comparison of algorithms across a range of values for the parameter $k$. Finally, we provide a new benchmark corpus of~$174$ networks drawn from~$6$ domains.
Our algorithms retain most of the performance of the state-of-the-art while reducing running time by orders of magnitude. Specifically, a meta-learner approach is on average only $20\%$ less effective than the state-of-the-art on held-out data, but about $75-130$ times faster. Further, the meta-learner's performance exceeds the state-of-the-art on about $20\%$ of networks, and the magnitude of its running time advantage is maintained on much larger networks.

\end{abstract}

\maketitle

\section{Introduction}

Influence maximization~\cite{kempe_influence_maximization_2003,chen2010scalable} is one of the most intensively studied problems in data mining, machine learning, and social network analysis. Given some model of information diffusion, commonly the \emph{independent cascade} of Kempe, Kleinberg, and Tardos~\cite{kempe_influence_maximization_2003}, the task is to determine where we should seed information such that it spreads as widely as possible. Formally, given a graph $G$ and
$S \subseteq V(G)$, for each $i \in V(G)$ let $\pi_i(S)$ denote the probability with which $i$ is activated by an independent cascade seeded at $S$; we will just write $\pi_i$ when the set $S$ is clear. Then for a given budget $k$, influence maximization is the problem of identifying a set $S \subseteq V(G)$ of cardinality $k$ which maximizes $\sum_{i \in V(G)} \pi_i$.

Influence maximization is commonly motivated by commercial advertising, e.g.,~\cite{domingos2001mining,richardson2002mining,chen2010scalable}
but is also relevant in various other applications, including
public health~\cite{leskovec2007cost,yadav2016using}, the distribution of economic opportunities~\cite{banerjee2013diffusion}, and the spread of scientific knowledge~\cite{rogers2014diffusion}.
In these applications, the desiderata include not only widespread but also \emph{fair} dissemination of information. The difficulty of the latter goal is a natural consequence of structural heterogeneity in networks. Highly-connected or centrally located individuals have more opportunities to receive and spread information~\cite{cheng2019,PICARD201877,toth2021}, while peripheral individuals with few connections participate less often in information exchanges~\cite{PICARD201877,Fish_2019}.
For example, in a pandemic, access to crucial resources---such as money, food and healthcare---is more difficult for socially disadvantaged groups, in part due to their more limited connectedness in social networks~\cite{GAROON20081133}. Similarly, connectedness shapes employment opportunities in professional social networks such as LinkedIn:\ well-connected job seekers are likely to fill lucrative openings sooner than others~\cite{weaktiesinlinkedin2022}.
This situation has motivated the study of several variations of influence maximization, for example to ensure equitable information access with respect to demographic groups~\cite{stoica2018algorithmic}.
In this paper, we focus on the formulation of Fish \emph{et al}.~\cite{Fish_2019}, which adopts a \emph{Rawlsian}~\cite{rawls2009theory} notion of fairness in which the goal is to improve the information access of the worst-off individuals. Formally, the problem is as follows:

\optproblemdef{\textsc{Maximin Influence Maximization}}{A graph $G = (V, E)$ and an integer $k$.}{Select $k$ vertices $S \subseteq V$ maximizing $\min_{i \in V}\pi_i$.}

\begin{figure}[t]
\centering
    \includegraphics[width=0.4\textwidth]{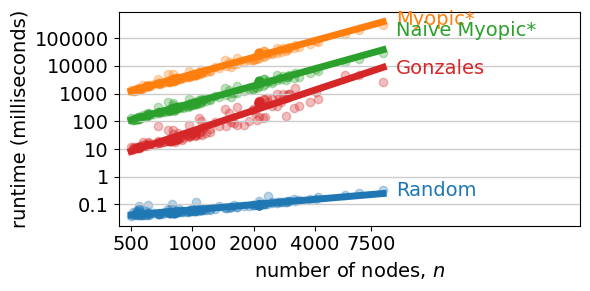}
    \caption{Algorithm runtime to select 10 new seeds vs.\ network size for  algorithms in~\cite{Fish_2019}, averaged over 10 runs on an introduced large set of networks (see Section~\ref{sec:corpus}). Algorithms requiring a Monte Carlo simulation (\texttt{ProbEst}) to select seeds are denoted by a $*$.}
    \label{fig:algos_timing_n_small}
\end{figure}

This maximin variant is \cclass{NP}-hard, and even a constant-factor approximation would imply \cclass{P} = \cclass{NP}~\cite{Fish_2019}, in stark contrast to the greedy $(1-1/e)$-approximation for the standard objective~\cite{kempe_influence_maximization_2003}. However, Fish~\emph{et al.} show that a heuristic approach (\texttt{Myopic}; see Section~\ref{sec:algos_prior_work}) optimizes the objective well in practice. Indeed, this approach even performs well when evaluated via the classic influence maximization objective, which it does not directly optimize. Unfortunately, this heuristic relies upon repeated
Monte Carlo simulations of the activation probabilities $\pi_i$, which are a severe computational bottleneck and thereby constrain the scalability of the method; see~\Cref{fig:algos_timing_n_small}.
Additionally, the method has only been evaluated on a small corpus of six networks.

We are thus motivated to (a) introduce new algorithms which alleviate or completely avoid the bottleneck imposed by probability estimation; (b) compare the performance of our methods against that of \texttt{Myopic}
in a principled manner,; and (c) expand the number and diversity of networks on which the
\textsc{Maximin Influence Maximization} problem has been studied.

\subsection{Related Work and Preliminaries}\label{sec:related-work}

\subsubsection{Fairness in ML and Social Networks.}
Fairness in machine learning is a well-studied area, and includes notions of fairness based both on individual characteristics and on group demographics \cite{feldman2015certifying, hardt2016equality, dwork2012fairness} (for surveys, see \cite{caton2024fairness, pessach2022review}). 
Recently, questions of fairness of information access of individuals and demographic groups in a network have come to the fore \cite{Fish_2019, stoica2019fairness, stoica2020seeding, ali2021fairness, mehrotra2022revisiting, tsang2019group-fairness, stoica2018algorithmic}. A significant body of work focuses on the probability that an individual receives information that spreads through a network. This includes studies of seeding information at nodes to improve individual~\cite{Fish_2019} or group~\cite{tsang2019group-fairness, stoica2020seeding} access; as well as interventions to add edges \cite{bashardoust2023reducing, becker2023improving, bhaskara2024optimizing} under varying notions of fairness. Other work considers notions of group fairness based on the network structure~\cite{jalali2023fairness, mehrotra2022revisiting, bashardoust2020information}. Saxena, Fletcher, and Pechenizkiy provide a recent survey of such work~\cite{saxena2024fairsna}.
Fair clustering of individuals has also received significant attention, e.g.,~\cite{chierichetti2017fair}.\looseness=-1

\subsubsection{Independent Cascade.}
Considerations about the access of individuals to resources in a network build on structural concerns about social networks pioneered by Granovetter~\cite{granovetter1973}. Necessary in any such study is a clear model for the dynamics of information propagation. 
Numerous models exist, notably including the independent cascade, generalized independent cascade, and linear threshold models~\cite{kempe_influence_maximization_2003,granovetter1978,goldenberg2001}.
A standard choice, also adopted in this paper, is the independent cascade,
which can be defined via an iterative spreading process. At each step~$i$, some subset $S_i$ of nodes is \emph{activated}, beginning with an initial seed set $S = S_0$. In round~$i+1$, each edge $uv$ with $u \in S_i$, $v \notin \bigcup_{j \leq i} S_j$ activates $v$ independently with probability $\alpha$. The process stops after the first round $i$ in which no nodes are activated, i.e., $S_i = \emptyset$, and a node~$v$ is said to be activated by the cascade if it is activated in any round, i.e., $v \in \bigcup S_i$. 

\subsubsection{Probability Estimation}
Important in any analysis of independent cascades is the ability to measure the probability $\pi_i$ with which node $i$ is activated. Unfortunately, exact computation of $\pi_i$ is 
\cclass{\#P}-hard~\cite{chen2010scalable}. The standard approach is to estimate these values via \emph{reverse influence sampling} (RIS)~\cite{borgs2014maximizing, tang2014influence}, which may also be thought of as performing a series of Monte Carlo simulations of the cascade process. Theoretical bounds on the number of simulations needed to satisfy a given error tolerance~$\varepsilon$ have a quadratic dependence on~$\varepsilon$ in addition to quasilinear dependence on network size~\cite{tang2014influence}, and so in practice it is common to fix a reasonable number $R$ of simulations, e.g., $R= 1000$. Algorithms for influence maximization and related problems then invoke a linear-time (regarding $R$ as a constant) subroutine, referred to here as \texttt{ProbEst}, to provide
estimated $\pi_i$ values. In practice, however, algorithms requiring such a subroutine can be orders of magnitude slower than those avoiding probability estimation altogether; we again refer to~\Cref{fig:algos_timing_n_small}. In the influence maximization literature, much effort has been expended to lessen the complexity of \texttt{ProbEst} while retaining quality guarantees, e.g.,~\cite{tang2015influence,nguyen2016stop,huang2017revisiting}, as well as to empirically compare various strategies~\cite{ohsaka2020solution}. In this work, our approach is to develop methods which avoid RIS entirely and evaluate the resulting solution quality via empirical comparison against the state-of-the-art. A purely heuristic approach, while difficult to accept for the classic influence maximization problem, is palatable in our setting because strong inapproximability bounds are known for \textsc{Maximin Influence Maximization} even in the presence of an oracle which perfectly computes the $\pi_i$ values~\cite{Fish_2019}.

\subsection{Summary of Contributions}\label{sec:contributions}

\paragraph{Network Corpus.} In~\Cref{sec:corpus}, we introduce a large and structurally diverse corpus of~$174$ networks on which to benchmark both our algorithms and the state-of-the-art. These networks are drawn from six domains: biological, social, economic, technological, transportation, and informational. Though our primary interest is in social networks, building a corpus from multiple domains enhances the structural diversity of our benchmark set; see the discussion in~\Cref{sec:corpus}. Where relevant, we report results by domain.

\paragraph{Spreadability.} When evaluating algorithms in the presence of an independent cascade, it is necessary to choose a value for the parameter $\alpha$ which governs the probability of information transmission along edges. It is common in the literature to
choose several values, under the assumption that small values fundamentally represent ``low''
information spread while large values represent ``high'' spread. However, this qualitative assessment of a particular $\alpha$ value is not consistent across networks, due to the varying underlying combinatorial structure. We introduce \emph{spreadability}, which enforces mathematical rigor in the assessment of a particular $\alpha$ value as ``low''- or ``high''-spreading for a given network. We use this framework to select appropriate $\alpha$ values for each network in our corpus, ensuring that we can measure algorithm performance under multiple distinct regimes of information spread.  

\paragraph{Performance Metric.} Even with a fixed $\alpha$ value and a single network, it is non-trivial to compare the performance of algorithms. Among other factors, this is because we must account for varying choices of budget $k$, random number generator seeding, and the impracticality of precisely measuring the problem objective. We introduce a performance metric which incorporates all of these factors, and allows us to evaluate the performance of an algorithm via a single number $\beta$. Intuitively, $\beta$ indicates the marginal improvement in the problem objective that we can expect if we use a given algorithm to add an additional information seed. We defer a more formal description to~\Cref{sec:algo_performance_metric}.  

\paragraph{Algorithms.} In~\Cref{sec:algos}, we introduce~$10$ new algorithms for the \textsc{Maximin Influence Maximization} problem, which we partition into three categories. The first two replace the Monte Carlo simulations used by the state-of-the-art to estimate activation probabilities with heuristics based on breadth-first search and personalized page rank. The complexity of our subroutines is similar to the Monte Carlo approach, but the hidden constants are vastly improved, and this is reflected in improved running times. The third category of algorithms eliminates probability estimation entirely, instead using topological features of the network to inform seed selection.\looseness=-1

\paragraph{Evaluation, Meta-Learning, and Scalability.}
In~\Cref{sec:experiments}, we conduct a comprehensive evaluation of our methods, together with a comparison against the state-of-the-art (\texttt{Myopic}~\cite{Fish_2019}).
In~\Cref{sec:algo_performance}, we show that on most networks at least one of our algorithms is nearly as effective as or even more effective than \texttt{Myopic}.
Meanwhile, our algorithms are much faster, improving on the running time of \texttt{Myopic} by $1$-$4$ orders of magnitude; see~\Cref{sec:algorithm-runtime}. We build on these insights by introducing a meta-learner (see~\Cref{sec:meta-learner}), which uses structural features of a network to predict which of our algorithms will be most effective on that network. This approach allows us to recover roughly $80\%$ of the performance of \texttt{Myopic} while reducing running time by factors of about $75 - 130$. Finally, in~\Cref{sec:large-networks} we show that these improvements in running time persist even on several much larger networks.

\begin{figure}
    \includegraphics[width=0.45\textwidth]{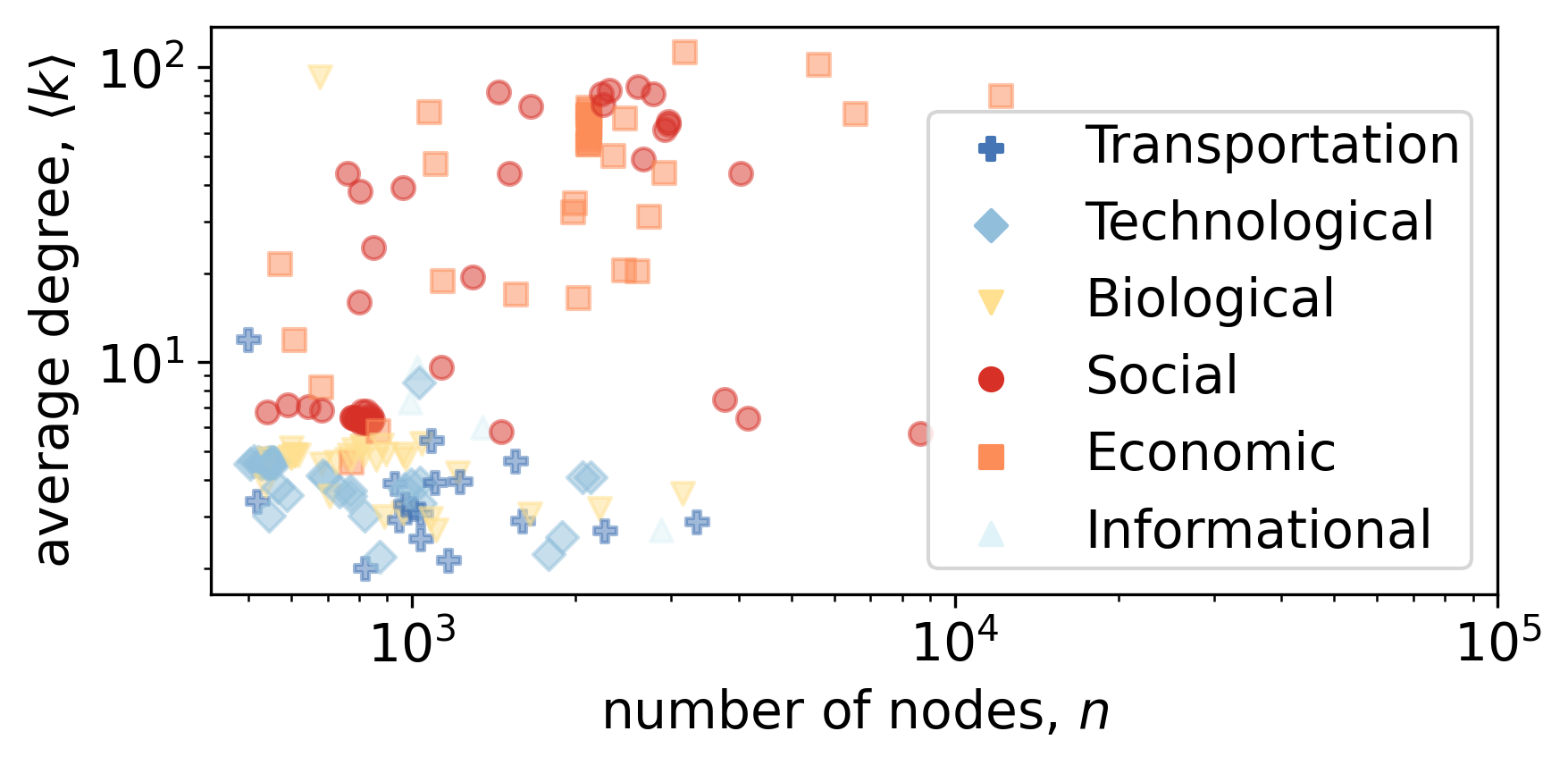}
    \caption{Average degree of a network as a function of network size (number of nodes) for the corpus of 174 networks from 6 distinct domains used in our study.}
    \label{fig:corpus_overview}
\end{figure}
\section{Network Corpus and Evaluation Methods}

\subsection{Network Corpus} \label{sec:corpus}

In order to investigate the effects of network structure on algorithm performance, we construct a corpus of $174$ networks from six domains: biological (34), social (44), economic (43), technological (32), transportation (17), and informational (4). We include non-social networks in our study in order to more fully characterize the behavior of algorithms on structurally diverse real-world networks. When relevant, we report our results by domain, so that social and non-social networks can be contrasted. An overview of the corpus is presented in~\Cref{fig:corpus_overview}, with summary statistics in~\Cref{table:corpus_stats}.

Networks were curated from the Index of Complex Networks~\cite{ICON}, a large-scale index of research-quality networks spanning all domains of science, as well as the
Netzschleuder network catalogue and repository~\cite{Netzschleuder} and the corpus of
Ghasemian, Hosseinmardi, and Clauset~\cite{Ghasemian_2019}. All networks included in our corpus are simple graphs, meaning their edges are undirected, unweighted, and there are no self-loops. Further, they are unipartite, i.e., they only have one type of node. 

Past work on such corpora indicates that domains (and even subdomains, e.g., online social networks vs.\ offline social networks) are highly distinguishable based on their structure alone~\cite{ikehara2017}. Hence, a specific effort was made to (i)~balance the classes, so that no domain was more than 25\% of the corpus, (ii)~avoid over-representing networks from particular sources (e.g., Twitter follower networks), and (iii)~ensure that the minimum network size was large enough to provide good results for information spreading tasks (minimum number of nodes $n_{\min}=500$). These choices improve the breadth and variety of network structure represented in the corpus, and its utility in the analyses of this study.

\begin{figure}[t!]
    \includegraphics[width=0.45\textwidth]{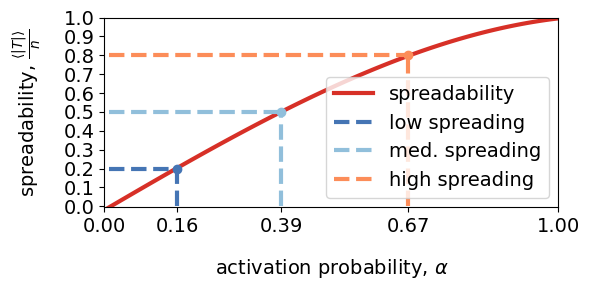}
    \caption{Spreadability on a network is quantified by the average fraction of a network's nodes $\langle |T| \rangle/n$ in a tree $T$ grown through an independent cascade from a random initial seed for a given $\alpha$. We define `low', `medium', and `high' spreadability as the $\alpha$ that activates, on average, $20\%$, $50\%$, and $80\%$ of the network, respectively.}
    \label{fig:spreadability}
\end{figure}

\subsection{Spreadability} \label{sec:ic}

\paragraph{Background.}
In our work, we leverage the independent cascade model~\cite{kempe_influence_maximization_2003} to study the spread of information in networks and estimate $\pi_i$. Given a network $G$ and a set of activated nodes $Q \subseteq V$, we grow a forest on $G$ under the independent cascade model by flipping a coin for each edge $e_{ij}$, where $i \in V \backslash Q$, $j \in Q$, exactly once, so that $i$ is added to $Q$ on a successful flip. Each edge $e_{ij}$ is considered as a transmission path at most once, and we define the probability of successful transmission to be $\alpha$. We note that mathematically, it is equivalent to conceptualize the independent cascade as follows: delete every edge in the graph independently with probability $(1 - \alpha)$; the set $Q$ of activated nodes is exactly the set of nodes that remain reachable from the seed set $S$. As noted earlier, computing the exact probability of activation for a node $i$, denoted $\pi_i$, is \textsc{\#P}-hard~\cite{chen2010scalable}. As such, we adopt the standard Monte Carlo simulation approach, but return later to consider practical consequences of this choice.

Briefly, using \texttt{ProbEst}, given transmission probability $\alpha$, number of simulation rounds $R$, and a seed set $S$, we estimate $\pi_i$ for every $i \in V$ using $R$ independent cascades originating from $S$~\cite{Fish_2019}. 
The worst-case time complexity of \texttt{ProbEst} is $\mathcal{O}(R(|S|+2m))$~\cite{Fish_2019}, where $m$ is the number of edges in $G$. In practice, the computational cost of \texttt{ProbEst} increases both as the seed set grows and as $\alpha$ increases, because both changes tend to increase the size of the induced information cascades. Past work used $R=1000$ as a balance between statistical accuracy for $\pi_i$ and computational cost, and we follow this precedent~\cite{Fish_2019}.
Here, \texttt{ProbEst} is used as a subroutine in some algorithms, as well as to evaluate the performance of algorithms by estimating the achieved minimum activation probability of a computed seed set.

\paragraph{Selecting $\alpha$ values.}
Prior work evaluated algorithm performance using transmission probabilities \mbox{$\alpha \in \{0.3,0.4,0.5\}$}~\cite{Fish_2019}. However, the resulting information cascades, and hence the associated access probabilities, are not a simple function of $\alpha$; they instead also depend on the network's structure. For example, denser, more connected networks contain many more paths by which information can spread than do sparse networks. Hence the same $\alpha$ will tend to produce larger cascades on the former, and smaller cascades on the latter. To control for these structure-induced differences in information cascades, we introduce the concept of \textit{spreadability}, which jointly accounts for the impact of network structure and transmission rate $\alpha$ on the sizes of information cascades.

\begin{figure}[t!]
    \includegraphics[width=0.45\textwidth]{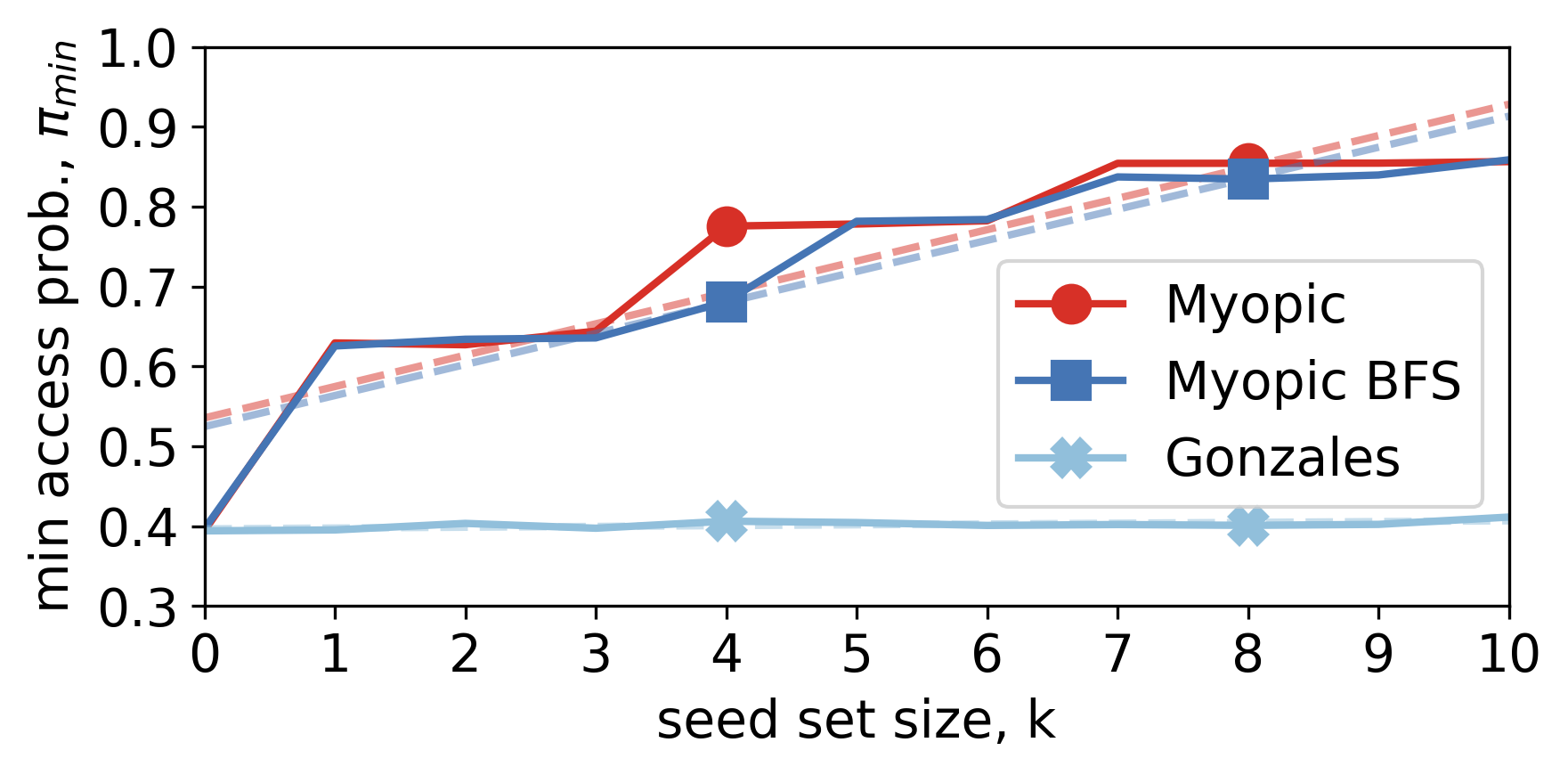}
    \caption{Minimum access probability $\pi_{\min}$ vs.\ seed set size $k$, 
    with a best fit line (\texttt{Myopic} $\hat{\beta}=0.039$), averaged over 20 runs, evaluated on a large economic network ($n=2113$ nodes, $m=57927$ edges), with $\alpha=0.4$ and a budget of $k=10$ seeds, plus one random initial seed. 
    }
    \label{fig:curves_with_slopes}
\end{figure}

We can construct a fine-grained spreadability function that relates a particular choice of $\alpha$ to the fraction of a network activated under the independent cascade model from a uniformly random initial seed. For a given $\alpha$, the spreadability $f(\alpha)$ on a particular network the fraction of nodes that are activated from a uniformly random initial seed, averaged over $R$ trials. This calculation produces a monotonically increasing curve, as seen in Figure~\ref{fig:spreadability}.
We find that computing spreadability for each \mbox{$\alpha \in \{0.01, 0.02,... 0.99\}$} provides ample resolution to choose $\alpha$ close to target spreadabilities of $0.2,0.5,0.8$ (which we refer to as ``low,'' ``medium,'' and ``high'' spreadabilities respectively). We set $R=1000$, as the spreadability curve tends to stabilize near this value and we get diminishing returns for larger $R$.

\subsection{New Metric for Algorithm Evaluation} \label{sec:algo_performance_metric}

As discussed in~\Cref{sec:contributions}, it is not straightforward to compare the performance of algorithms, even with a fixed network and $\alpha$ value. There are several reasons for this. First, we do not want to make strong assumptions about the ``most useful'' value for the parameter $k$ (the number of seeds to be added), and it is possible that algorithm $\mathcal{A}$ obtains a better objective score (the minimum activation probability $\pi_{\text{min}}$) than algorithm $\mathcal{A}'$ for one value of $k$, but not for another. Moreover, many algorithms (in particular those using \texttt{ProbEst}) have some randomness.
Finally, even our analysis of the quality of a solution is inexact; recall that exact computation of the $\pi_\text{min}$ achieved by a seed set $S$ is \cclass{\#P}-hard~\cite{chen2010scalable}. Thus, our ability to compare two solutions is limited by the precision of \texttt{ProbEst}. 

We overcome these challenges by introducing a metric $\beta$, which intuitively measures the marginal gain in the problem objective (the minimum activation probability $\pi_{\text{min}}$) which we can expect when asking an algorithm to add one additional seed.
Formally, for a given algorithm $\mathcal{A}$, network $G$, independent cascade parameter $\alpha$, and budget $k$, we run $\mathcal{A}$ on $(G, \alpha, k)$ and record a value $\pi_\text{min}(k)$ indicating the achieved objective score. We repeat for each $k \in \{1, 2, \ldots, 10\}$, computing the points $(k, \pi_\text{min}(k)) \in \mathbb{N}\times [0, 1]$.
We then record the slope $\beta_1$ of the line of best fit for these points.
We repeat the entire process multiple (generally~$20$) times, producing slopes $\beta_1, \beta_2, \ldots$. The metric $\beta$ is the mean of these values, i.e., the average slope of the line of best fit; see~\Cref{fig:curves_with_slopes}.
Henceforth, when comparing the performance of algorithms, we do so via the metric $\beta$.\looseness=-1

\section{Algorithms for Fair Information Access}\label{sec:algos}

\subsection{Algorithms from Prior Work} \label{sec:algos_prior_work}

Four previously introduced heuristics for choosing seed nodes to maximize $\pi_{\rm min}$ are \texttt{Greedy}, \texttt{Myopic}, \texttt{Naive Myopic} and \texttt{Gonzalez}, along with \texttt{Random} as a baseline for comparison~\cite{Fish_2019}. 
\texttt{Random} selects $k$ seeds by uniformly sampling nodes from the network. Given a partial seed set, \texttt{Gonzalez} selects as the next seed the node that is the furthest (in the shortest-path metric) from all nodes in the current set~\cite{Fish_2019}. The \texttt{Greedy} algorithm iteratively selects a new seed by choosing the node with the highest marginal gain relative to the current seed set according to \texttt{ProbEst}. Due to its immensely high computational cost, \texttt{Greedy} is not practical for most networks~\cite{Fish_2019}, and is not used in our study.

In contrast, \texttt{Myopic} uses \texttt{ProbEst} with the current seed set, and selects as the next seed the node with the lowest $\pi_i$. \texttt{Naive Myopic} is similar to \texttt{Myopic}, but only runs \texttt{ProbEst} once, at initialization, and then selects as the seed set the $k$ nodes with the lowest $\pi_i$ values. In past work, on a small set of networks, \texttt{Myopic} was found to perform best. However, because \texttt{Myopic} depends on \texttt{ProbEst}, it is computationally expensive, which limits its applicability to large networks. In practice, Myopic is always the second slowest algorithm, after \texttt{Greedy}~\cite{Fish_2019}.

\begin{figure}[t!]
    \centering
    \includegraphics[width=0.35\textwidth]{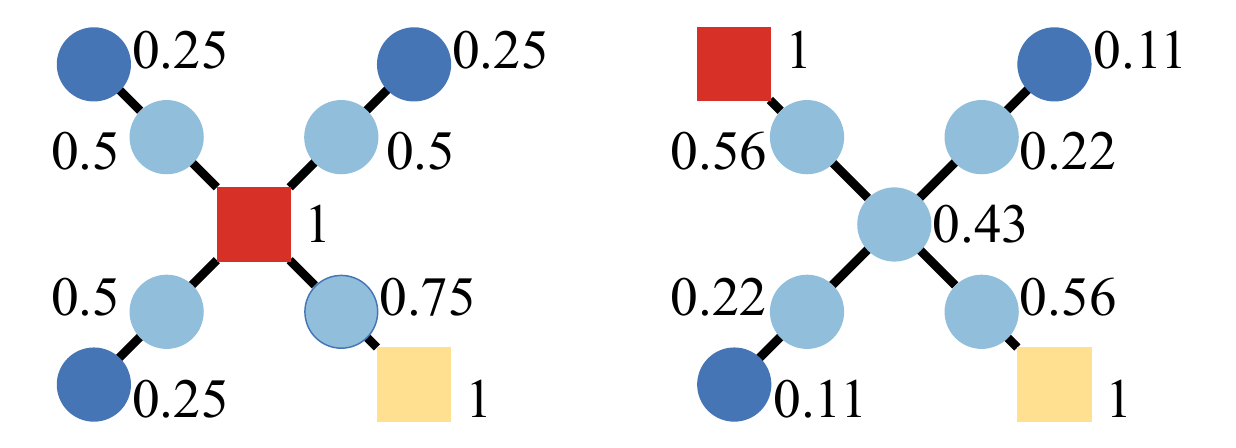}
    \caption{Illustrations of two runs of \texttt{Myopic} for different initial seeds (red), with new selected seeds (yellow), and fixed $\alpha=0.5$. Numbers indicate $\pi$ for each node after the new seed is selected. Initialization significantly affects the performance of \texttt{Myopic}.}
    \label{fig:seeding_methodology}
\end{figure}
\paragraph{Algorithm Initialization}

\begin{figure*}
    \includegraphics[width=1\textwidth]{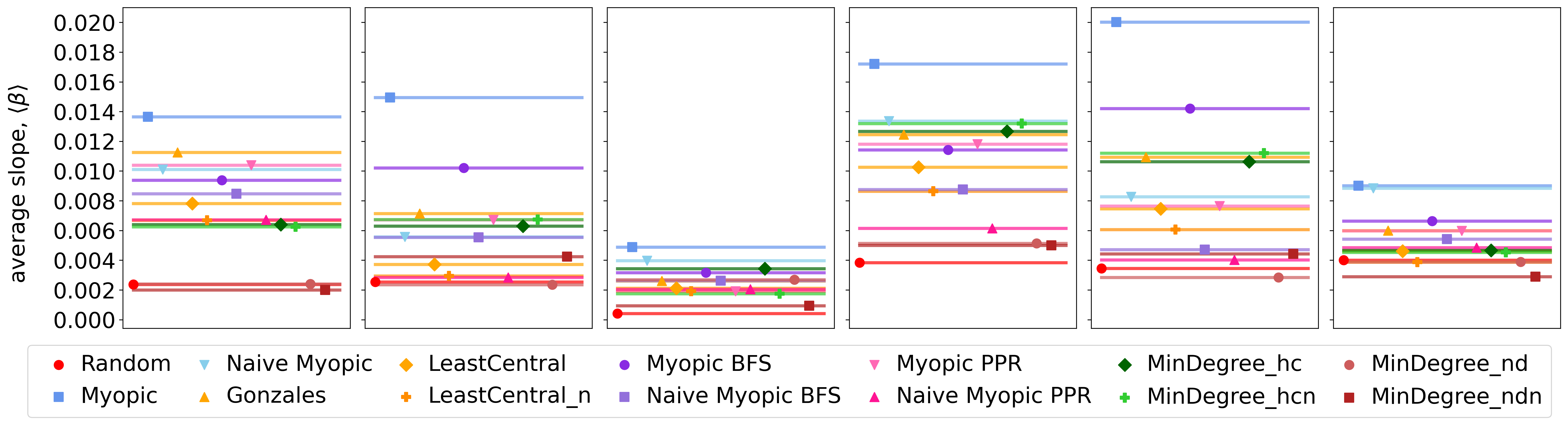}
    \caption{Mean performance of the intervention algorithms on each domain in the corpus under medium spreadability. Each algorithm's performance is averaged over a given domain, 20 runs per network.}

    \label{fig:algo_mean_performance}
\end{figure*}
\texttt{Myopic}, \texttt{Naive Myopic}, and \texttt{Gonzalez} all start with an 
initial seed. 
Past work chose this initial seed to be the highest degree node. 
Our initial experiments indicate that this choice confers a substantial 
advantage to these methods (\Cref{fig:seeding_methodology}), and that some of the previous positive results 
are thus attributable to this initial seed choice rather than to the algorithms' subsequent choices.
To mitigate this bias 
we instead initialize all heuristics 
with a seed set composed of a single uniformly randomly selected initial seed. Moreover, for each evaluation round on a particular network, we initialize all algorithms to use the same random seed, which further controls for the effects of different initial seed choices. This initial seed choice is not counted against the budget $k$ (\Cref{fig:curves_with_slopes}).

\subsection{New Fast Algorithms}

In addition to the four heuristics from prior work, here we introduce 10 new heuristics to maximize the minimum $\pi_i$ on a budget. These heuristics are designed to be computationally lightweight, scaling to far larger networks, 
while also matching or exceeding the performance of \texttt{Myopic} on the corpus. We group the new algorithms into three families: BFS-based, PPR-based and Topology-based.

\paragraph{BFS-based: Myopic BFS and Naive Myopic BFS}
In the two BFS-based heuristics, \texttt{Myopic BFS} and \texttt{Naive Myopic BFS}, we swap the \texttt{ProbEst} component of \texttt{Myopic} and \texttt{Naive Myopic} with a simple breadth-first search to estimate $\pi_i$. The breadth-first search component is initialized with a random seed $k_0$, and transmission probability $\alpha$. It proceeds to ``peel'' the network, starting at $k_0$, in breadth-first fashion, estimating $\pi_i$ for each node as the probability that $i$ receives a transmission from $k_0$ through any nodes it connects to in the previous BFS layer, as well as through nodes it is connected to in its own layer. All subsequent iterations update existing $\pi_i$ estimates during the breadth-first traversal from new candidate seeds.\looseness=-1

While this approach does not exactly measure $\pi_i$, it is much faster than \texttt{ProbEst}, taking $\mathcal{O}(n \cdot \langle k \rangle)$ per iteration, because for most networks, including those in our corpus, the mean degree $\langle k \rangle \ll 1000$ (\Cref{fig:corpus_overview}). The key design principle of this algorithm is to capture complex network structures that \texttt{Gonzalez} could not account for. 
In Figure~\ref{fig:curves_with_slopes}, we find that \texttt{Myopic BFS} almost matches \texttt{Myopic}'s performance in situations where \texttt{Gonzalez} lags behind.

\paragraph{PPR-based: Myopic PPR and Naive Myopic PPR}
In the PPR-based heuristics,
\texttt{Myopic PPR} and \texttt{Naive Myopic PPR}, 
we use 
Personalized PageRank (PPR) to estimate $\pi_i$ instead of \texttt{ProbEst} or BFS. Personalized PageRank (implemented with \textsf{networkx} \cite{SciPyProceedings_11}) performs a random walk that probabilistically restarts from nodes in the seed set~\cite{PPR1998}. 
The ranking produced by PPR is not a direct 
estimate of $\pi_i$; we treat the PPR values as being correlated, such that a lower PPR score is a proxy for a lower $\pi_i$ value.

\paragraph{Topology-based: LeastCentral and MinDegree.} These heuristics
are based on the intuition that nodes with low $\pi_i$ values have distinctive structural positions or patterns of connectivity. \texttt{LeastCentral} selects the non-seed node $i$ with the lowest closeness centrality $c_i$ as the next seed. Similarly, \texttt{LeastCentral\_n} 
selects the lowest centrality node $i$'s highest degree neighbor as the next seed. Here, the closeness centrality of a node $i$ is the inverse of the average shortest path length from $i$ to all other reachable nodes in the same connected component~\cite{FREEMAN1978215,newman18}. Lower closeness centrality implies the node is less reachable by the rest of the network, and therefore is expected to have lower $\pi_i$.

The four remaining topology-based heuristics exploit a network's degree structure to make decisions, based on 
the observation that in practice \texttt{Myopic} tends to select nodes with low degree as seeds. The first two heuristics take a non-seed node with the lowest degree, breaking ties by choosing the node with the lowest harmonic centrality~\cite{boldi2013axioms,newman18} among same-degree nodes. In the case of \texttt{MinDegree\_hc}, it chooses that node itself as the next seed, while \texttt{MinDegree\_hcn} chooses the highest-degree neighbor of that node. The two remaining variations replace harmonic centrality in the aforementioned logic with the neighbor degree, i.e. breaking ties by choosing the node with the highest neighbor degree (sum of degrees across neighbors). These heuristics are called \texttt{MinDegree\_nd} and \texttt{MinDegree\_ndn}, respectively.

\section{Experimental Results}\label{sec:experiments}
\begin{figure*}[t!]


    \includegraphics[width=0.9\textwidth]{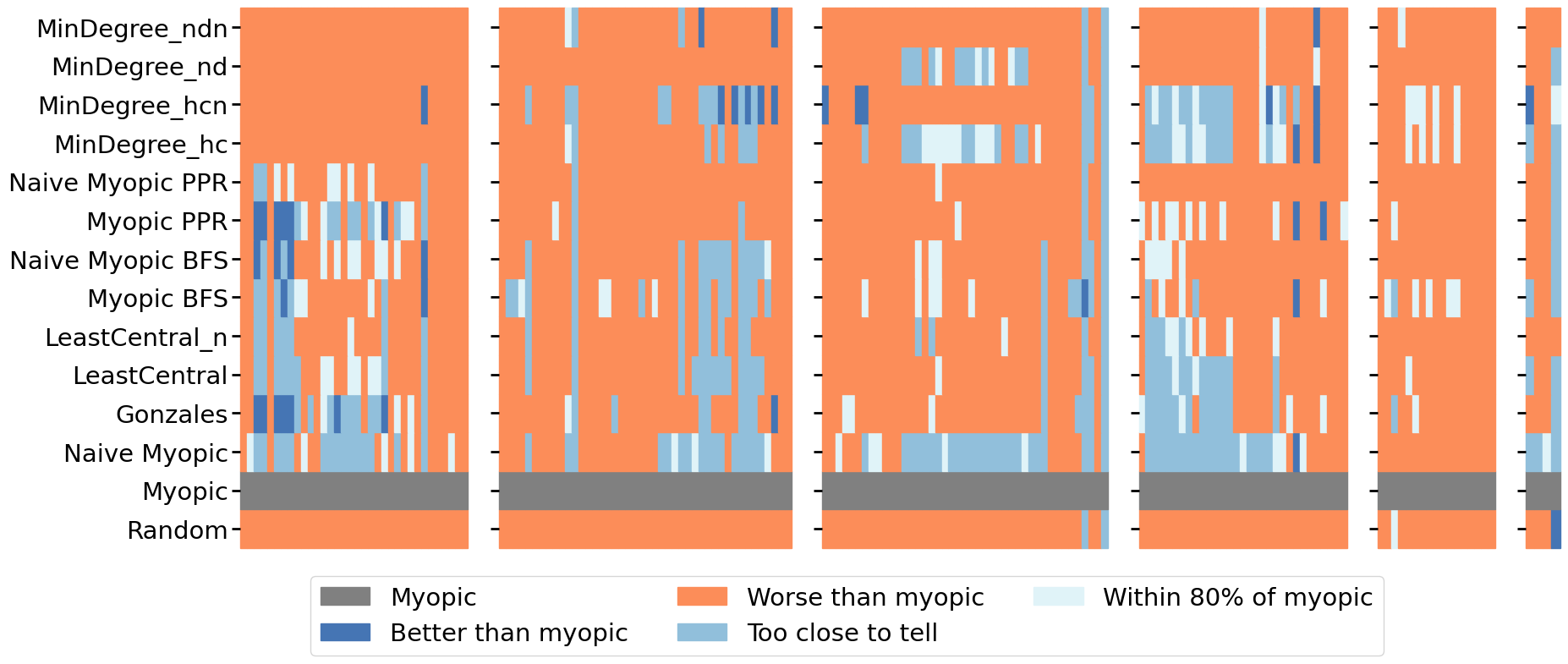}


    \caption{Performance of intervention algorithms on the network corpus, relative to \texttt{Myopic} and sorted in ascending order by network size within each domain. 
    Under medium spreadability, $24\%$ of networks have no algorithm better than or within $80\%$ of \texttt{Myopic}'s performance. ``Equivalent'' is defined as within one standard error of $\beta$ for \texttt{Myopic}; typically about 0.001.}
    \label{fig:algo_performance_heatmaps}
\end{figure*}

\subsection{Performance of Algorithms on the Corpus} \label{sec:algo_performance}

We evaluate and compare the performance of 14 algorithms in total (10 new algorithms and 4 from prior work), applied to all $174$ networks in the corpus.
Our code and data are available at \url{https://github.com/TheoryInPractice/FairInfoAccessHeuristics}.
We are interested in algorithms that operate well on a tight budget and so let $k \in \{1, 2, \ldots, 10\}$ seed nodes. For each spreadability level (low, medium, high), we produce a $10 \times 14 \times 174$ matrix, where each entry is the $\beta$ performance of an algorithm after adding $k$ seeds in a network, averaged over $20$ runs. We focus on the medium spreadability regime here, and include results for low and high spreadability in~\Cref{appendix:supplementary-results}.
Figure~\ref{fig:algo_mean_performance} displays the mean performance of each algorithm on networks by domain under medium spreadability (see~\Cref{fig:algo_mean_performance_high} for low and high spreadability results). Across domains, \texttt{Myopic} produces the best average performance.

To further examine the results, 
we sort within each domain by network size in ascending order, and then score algorithms as better than, ``equivalent'' to (within one std.\ err.), within 80\% of, or worse than \texttt{Myopic} (see~\Cref{fig:algo_performance_heatmaps}; low and high spreadability results in~\Cref{fig:supp_algo_performance_heatmaps}). In this way, we can assess whether the better average performance of \texttt{Myopic} applies to individual networks compared to other algorithms.
This experiment reveals that \texttt{Myopic} is not universally the best algorithm for all networks in any spreadability setting; on many individual networks other algorithms perform equivalently or better. 
In the medium spreadability setting, for only $24\%$ of networks is there no algorithm that performs at least $80\%$ as well as \texttt{Myopic} (\Cref{fig:algo_performance_heatmaps}).

\begin{figure}[h!]
\centering
    \includegraphics[width=0.40\textwidth]{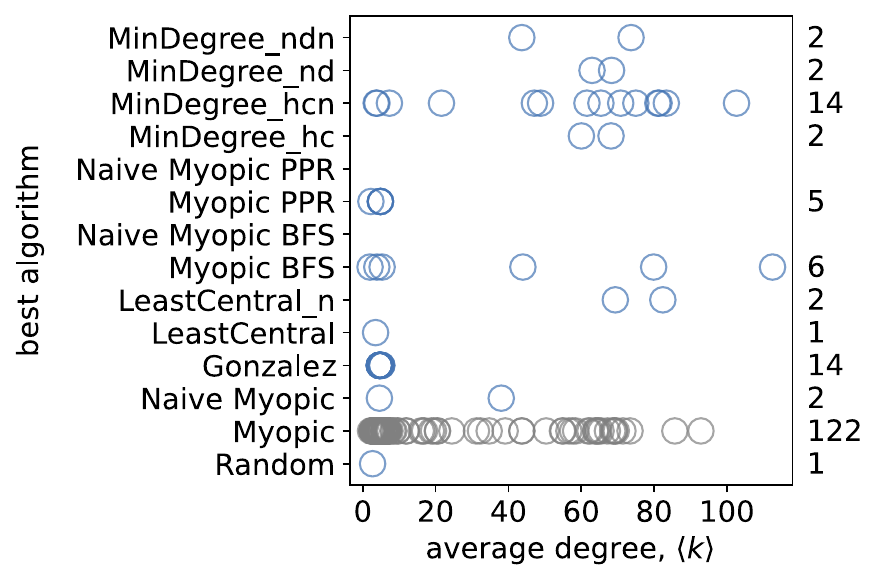}
    \caption{Best-performing algorithm vs.\ mean degree of the network (medium spreadability), for all networks. Counts on the right show total circles per line, i.e., the number of times an algorithm was the best over the whole corpus.}
    \label{fig:best_vs_avg_degree}
\end{figure}

This variability in performance across networks suggests that network structure plays a critical role in governing the relative performance of different algorithms. Figure~\ref{fig:best_vs_avg_degree} plots the instances for which each algorithm was the best performing algorithm on a network against that network's mean degree $\langle k \rangle$. \texttt{Myopic} tends to perform better on networks with lower average degree, although it also performs well on many networks with larger mean degree. 
In practice, we find that the final seed set produced by \texttt{Myopic} is often composed primarily of low-degree nodes located in a network's periphery, and these nodes may be too far removed from other disadvantaged nodes to meaningfully improve their $\pi_i$ values.

A second takeaway is that a few specific algorithms tend to perform better than \texttt{Myopic} in certain settings (\Cref{fig:best_vs_avg_degree}), specifically \texttt{MinDegree\_hcn} and \texttt{Gonzalez}. The performance of \texttt{MinDegree\_hcn} in particular tends to improve over \texttt{Myopic} with increasing average degree, while \texttt{Gonzalez} does best in networks with very low mean degrees. Furthermore, we note that in Figure~\ref{fig:algo_mean_performance_low}, on average \texttt{MinDegree\_hcn} outperforms \texttt{Myopic} in the economic domain, and from Table~\ref{table:corpus_stats}, we see that the economic domain has the highest mean degrees.
The \texttt{MinDegree\_hcn} algorithm selects as seeds the highest-degree neighbors of low-degree nodes. 
As result, new information cascades seeded at these nodes will tend to spread quickly to a number of disadvantaged nodes in the network.

\subsection{Algorithm Runtime}\label{sec:algorithm-runtime}
We evaluate all algorithm runtimes on the corpus, selecting $k\!=\!10$ new seeds, averaged over $10$ runs. For fair comparison, we run all algorithms on a single core of an AMD Ryzen 5900X, overclocked to $5.00$Ghz, with 32GB RAM, and measure the runtime in milliseconds. Two major performance bottlenecks are \texttt{ProbEst}, used by \texttt{Myopic} and \texttt{Naive Myopic}, and an All-Pairs-Shortest-Paths (APSP) computation, used by \texttt{Gonzalez}, \texttt{LeastCentral}, \texttt{LeastCentral\_n}, \texttt{MinDegree\_hc}, and \texttt{MinDegree\_hcn}. Both \texttt{ProbEst} and \texttt{APSP} have efficient parallel implementations, but we restrict them to a single core to ensure fair comparison with other algorithms.

\begin{figure}
    \includegraphics[width=0.45\textwidth]{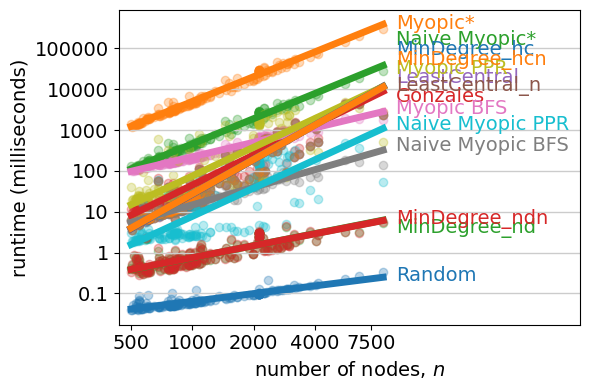}
    \caption{Runtime of old and new algorithms across the network corpus with $k=10$ seeds averaged over 10 runs, showing a substantial advantage in running time for the new algorithms. Note: algorithms with * require \texttt{ProbEst}.}
    \label{fig:algos_timing_n}
\end{figure}

All of the new algorithms are substantially faster than \texttt{Myopic} and \texttt{Naive Myopic} (\Cref{fig:algos_timing_n}). Because they only require sorting two lists, \texttt{MinDegree\_ndn} and \texttt{MinDegree\_nd} are the most efficient, improving running times over \texttt{ProbEst}-based algorithms by a factor of 1000-10000x, depending on the size of the network (\Cref{fig:algos_timing_n}). \texttt{BFS}-based algorithms are marginally slower than these fastest algorithms, and algorithms that use an All-Pairs-Shortest-Paths (APSP) calculation fall between the \texttt{ProbEst} and \texttt{BFS} algorithm groups. As expected from the asymptotics, \texttt{BFS} algorithms tend to scale more slowly (in terms of runtime) with network size than do \texttt{ProbEst} algorithms, while APSP algorithms scale more quickly (\Cref{fig:algos_timing_n}).

The low upfront cost of APSP-based algorithms makes them far faster than \texttt{ProbEst}-based algorithms in many practical settings, being 10-100x faster on networks with less than $n=10^6$. However, the asymptotic cost of APSP is cubic in the network size, implying that for sufficiently large networks, \texttt{ProbEst} will be faster. For our corpus, we estimate the crossover point when Myopic becomes more efficient than \texttt{MinDegree\_hc} to occur between $1,262,000$ and $1,515,000$ nodes (95\% CI, from 1000 bootstraps). 
However, approximation algorithms for APSP could potentially extend their practical efficiency much further.

\subsection{A Fast Meta-Learning Algorithm}\label{sec:meta-learner}

We can exploit the variability in algorithm performance, and the fact that even among non-\texttt{Myopic} algorithms no alternative is superior on all networks, by introducing a meta-learner algorithm that combines multiple scalable heuristics to approximate the state-of-the-art performance of the \texttt{Myopic} algorithm. We compare this algorithm to a fast ensemble algorithm that uses an oracle to make perfect predictions about which scalable algorithm is best to apply on a particular test network, thus providing an upper bound on the meta-learner's possible performance. 

The meta-learner algorithm leverages the scalability of non-\texttt{ProbEst} algorithms while retaining good overall performance under all spreadability regimes. The task is as follows:\ given a particular network $G$ and knowledge of the information's spreadability (low, medium, or high), select the scalable algorithm with the best marginal benefit $\beta$ for improving information access.

As shown previously, many of the heuristics do not perform well, and so we begin by narrowing the set of available algorithms. For each of the three spreadability settings, we select the set $\mathcal{X}$ of five algorithms (excluding \texttt{Myopic} and \texttt{Naive Myopic}) that maximize the number of networks for which at least one among the set performs at least $80\%$ as well as \texttt{Myopic}. This produces sets 
\begin{itemize}
    \itemsep-0.1pt
    \item $\mathcal{X}_{\rm high}\!=\!\{$\texttt{Gonzales}, \texttt{LeastCentral}, \texttt{Myopic} \texttt{BFS}, \\ ${}^{}$\hspace{10mm} \texttt{Naive Myopic} \texttt{BFS}, \texttt{MinDegree\_hc}$\}$
   \item $\mathcal{X}_{\rm medium}\!=\!\{$\texttt{Gonzales}, \texttt{Myopic} \texttt{BFS}, \texttt{Myopic} \texttt{PPR}, \\ ${}^{}$\hspace{14mm} \texttt{MinDegree\_hc}, \texttt{MinDegree\_hcn}$\}$
    \item $\mathcal{X}_{\rm low}\!=\!\{$\texttt{Gonzalez}, \texttt{Myopic} \texttt{BFS}, \texttt{Myopic} \texttt{PPR}, \\ ${}^{}$\hspace{9mm} \texttt{MinDegree\_hcn}, \texttt{MinDegree\_ndn}$\}$
\end{itemize}

For the meta-learner algorithm, we then learn a 5-way random forest classifier to predict the best algorithm in $\mathcal{X}$ to apply to a given network, using nine of the network's topological features as the feature set (\Cref{fig:ml_importances}). We train and evaluate the meta-learner approach using an 80-20 train-test split among networks in the corpus, with meta-learner algorithm selection and model training both performed on the training set, and we report the mean performance over the test set. The meta-learner's runtime is the runtime of the trained model and the single algorithm it selects.

In the fast ensemble algorithm, for a given network in the corpus, the oracle runs all algorithms in $\mathcal{X}$ and evaluates the performance of each using \texttt{ProbEst} to calculate their respective $\beta$s, and then returns the single algorithm with the highest $\beta$ for that network. In this way, the oracle acts like an optimal classifier over $\mathcal{X}$ (cf.\ the meta-learner algorithm). The runtime of the fast ensemble algorithm is simply that of the single algorithm it selects.

\begin{figure}[t!]
\includegraphics[width=0.45\textwidth]{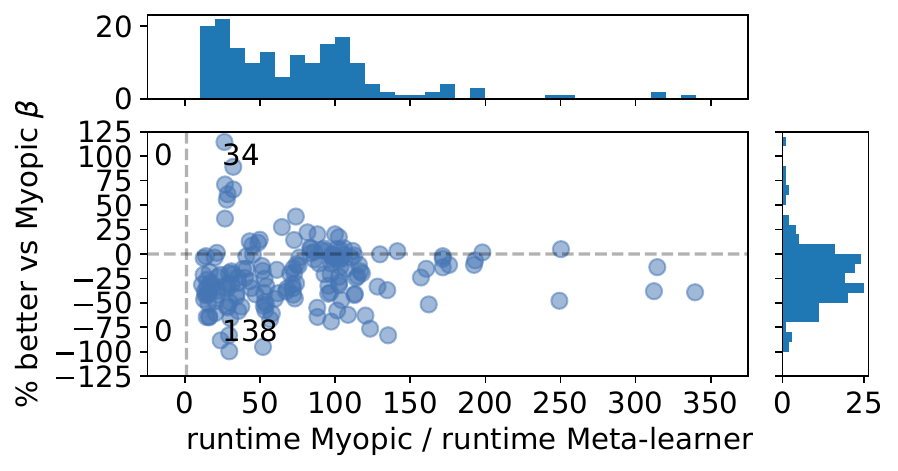}
    \caption{Performance difference vs.\ speedup for the meta-learner algorithm under medium spreadability, with marginal histograms, averaged over $1000$ runs. Extreme outliers have been removed for visualization purposes. Average performance difference relative to \texttt{Myopic} is $-20.11 \%\pm 29.34$ (mean $\pm$ stddev), for an average speedup factor of $76.28 \pm 64.07$. For 34 of the networks (19.8\%) the meta-learner strictly outperforms \texttt{Myopic}.}
\label{fig:metalearner_med}
\end{figure}

Compared to \texttt{Myopic}, the meta-learner is dramatically more efficient, with an average runtime that is $76.26\pm 64.07$ times faster under medium spreadability (\Cref{fig:metalearner_med}) and $133.35\pm 79.32$ times faster under high spreadability (\Cref{fig:metalearner_high}).  Improvement in scalability comes with a modest cost to performance, such that the meta-learner produces $\beta$ values that are, on average, $20.11\% \pm 29.34$ lower than those of \texttt{Myopic} under medium spreadability (with similar results for high); we note that the wide variance in these numbers reflects the broad range of difficulty across networks in the corpus. In contrast, the fast ensemble algorithm's performance is only $9.34 \% \pm 28.34$ lower for medium spreadability (similar results for high), indicating both room for improvement by the meta-learner with a better feature set as well as an upper limit to that improvement with the current scalable algorithms. We note, however, that lower performance is not universal:\ for 34 and 22 networks (20\% and 12.8\%), the meta-learner outperforms \texttt{Myopic} on medium and high spreadability, respectively (Figs.~\ref{fig:metalearner_med} and~\ref{fig:metalearner_high}). Under low spreadability, the fast meta-learner's average performance generally exceeds \texttt{Myopic} because of the inherent precision limitation of \texttt{ProbEst} in this setting.\looseness=-1

\subsection{Scaling to Larger Networks}\label{sec:large-networks}
In this section, we verify that our methods remain much faster than \texttt{Myopic} even on much larger networks. We consider two networks,
Email Network (EU Research Institution) (which we shorten to Email (EU)) and Google+ (2013). These have $\sim{34}$ thousand and $\sim{87}$ thousand nodes, respectively; other summary statistics are reported in~\Cref{table:large-networks}. On these networks, we evaluate \texttt{Myopic} against the five algorithms in $\mathcal{X}_\text{high}$, as defined in~\Cref{sec:meta-learner}: \texttt{Gonzales}, \texttt{Least Central}, \texttt{Myopic BFS}, \texttt{Naive Myopic}, \texttt{BFS}, and \texttt{MinDegree\_hc}.
We use high-spreadability $\alpha$ values. We ran these experiments on identical hardware equipped with 40 physical cores (Intel(R) Xeon(R) Gold 6230 CPU @ 2.10 GHz) and 19100 MB of RAM. As with the smaller networks, the budgets tested are $k \in \{1, 2, \ldots, 10\}$.

We change several parameters from our previous runs so that our experiments are better suited to the size of these networks. After initial experiments with the number of Monte Carlo simulations in $\texttt{ProbEst}$ set to $R=1000$, it became clear that 1000 simulation rounds was not enough for networks with tens of thousands of nodes, as some of our evaluations produced a negative $\beta$. With exact probability computations, $\pi_{\text{min}}$ would monotonically increase with each seed added, so $\beta < 0$ indicates highly inaccurate estimations. Thus, we increase the number of Monte Carlo simulations to $R = 10000$ in order to better estimate $\pi_i$ for large networks. However, due to the increase in simulation rounds and network size, running $\texttt{ProbEst}$ to evaluate the algorithms' performances is extremely expensive. To balance between computation time and a thorough analysis, we run only 3 trials of our algorithms with high-spreadability $\alpha$ values. We also use only $1000$ simulations when executing the spreadability computations (recall~\Cref{sec:ic}) to select the appropriate $\alpha$. We note that calculating the target $\alpha$ would have required approximately 30 days of compute time for Google+ (2013) with $R=10000$. After using $1000$ simulations to complete the spreadability computations, we tested the selected $\alpha$ values with $R=10000$, verifying that they activate on average $78.9\%$ and $79.4\%$ of Email (EU) and Google+ (2013), respectively, which is quite close to the target activation percentage of $80\%$.

\begin{figure}[t!]
\includegraphics[width=.45\textwidth]{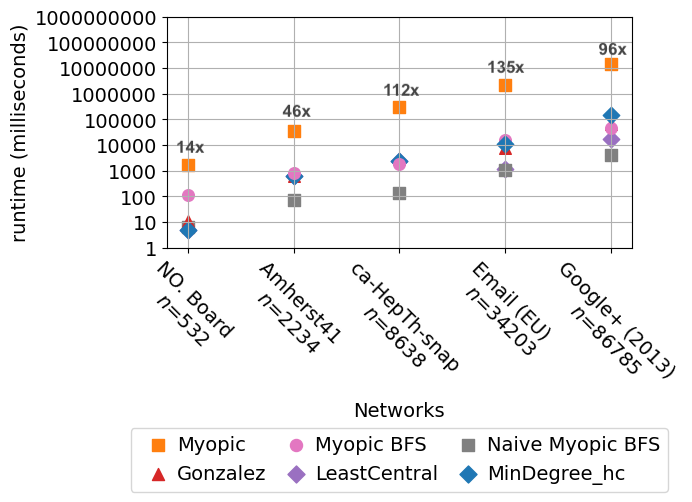}
    \caption{Runtime of \texttt{Myopic} and the algorithms of $\mathcal{X}_{\rm high}$ on five social networks of various sizes under high spreadability: Norwegian Board of Directors (2006), Amherst41, ca-HepTh-snap, Email (EU), and Google + (2013). Numbers in the plot indicate how much slower~\texttt{Myopic} is than the second-slowest algorithm for that network. }
\label{fig:larger_networks_timings}
\end{figure}

As with the broader corpus, when evaluated on the two larger networks, the five representative algorithms from $\mathcal{X}_{\text{high}}$ are orders of magnitude faster than $\texttt{Myopic}$, with some of the algorithms maintaining seed choice quality comparable to $\texttt{Myopic}$. Indeed, the \emph{slowest} of the algorithms from
$\mathcal{X}_\text{high}$ is~$135$ times faster than \texttt{Myopic} on Email (EU), and~$96$ times faster on Google+ (2013). These results (see~\Cref{fig:larger_networks_timings}) indicate that the running time advantage of our methods over \texttt{Myopic} is not diminished on networks much larger than those in the corpus of~\Cref{sec:corpus}.
Moreover, the performance loss of our methods (relative to \texttt{Myopic}) is similar (though slightly greater) to that observed for the larger corpus; see~\Cref{fig:larger_networks_performance} in~\Cref{appendix:large_networks}.\looseness=-1

\section{Discussion and Conclusion}

We evaluate new and existing algorithms, and introduce a meta-learning method for choosing $k$ seed nodes to maximize the minimum access probability $\pi_i$ of a node in a network. The meta-learner achieves a large (75x) speedup over the existing state-of-the-art, with a modest decrease in performance over a large corpus of networks. The previous Monte Carlo method for calculating 
a node's information access has high computational costs that limit the applicability of algorithms that rely on it, while the algorithms we introduced here scale-up to much larger networks.

To evaluate the algorithms, we introduce a large cross-domain network corpus, and a new performance metric $\beta$. We also introduce the concept of spreadability on a network to choose the independent cascade activation probability $\alpha$, accounting for the influence of a network's structural features on information spread.

Our findings have implications for the design of intervention strategies that aim to improve information access of the disadvantaged individuals in a network. This work suggests that the structure of the network plays a significant role in the performance of the algorithms, with average degree being a particularly important factor. The introduction of an ensemble method based on fast algorithms that do not rely on Monte Carlo simulation also leaves open the incorporation of future lightweight algorithms.

\bigskip

\section*{Acknowledgements}
The authors thank Daniel B.~Larremore and Zachary Kilpatrick for helpful feedback, and they acknowledge the BioFrontiers IT group at the University of Colorado Boulder for their support with data storage infrastructure, data management services, and High Performance Computing resources.\\

\section*{Funding}
This work was supported in part by National Science Foundation Awards IIS 1956183 (DRW, CJW, AC), IIS 1956286 (ACr, MW, FS, BS), and IIS 1955321 (SF).

\bibliographystyle{plain}
\bibliography{bibliography}

\clearpage

\renewcommand{\thefigure}{S\arabic{figure}}
\setcounter{figure}{0}
\renewcommand{\thetable}{S\arabic{table}}
\setcounter{table}{0}

\appendix
\begin{table*}
    \begin{tabularx}{\linewidth}{l|Y|Y|Y|Y|Y|Y|Y}
    & net.\ count & $\langle n\rangle$ & $\langle m\rangle$ & $\langle k \rangle$ & $\langle C \rangle$ & $\langle \ell_{\rm max} \rangle$ & $\langle \sigma^2 \rangle$ \\
    \hline
    full corpus & 174 & 1495.42 & 26288.68 & 22.60 & 0.24 & 13.79 & 1997.12 \\
    \hline
    biological & \hspace{0.5em}34 & \hspace{0.5em}927.50 & \hspace{0.5em}2830.24 & \hspace{0.5em}7.00 & 0.08 & 11.59 & \hspace{0.5em}285.87\\
    social & \hspace{0.5em}44 & 1647.75 & 29055.57 & 28.11 & 0.42 & 11.02 & 1099.06\\
    economic & \hspace{0.5em}43 & 2367.40 & 71946.23 & 51.97 & 0.38 & \hspace{0.5em}7.19 & 6647.82\\
    technological & \hspace{0.5em}32 & \hspace{0.5em}843.13 & \hspace{0.5em}1651.50 & \hspace{0.5em}4.07 & 0.07 & 17.84 & \hspace{1em}69.55\\
    transportation & \hspace{0.5em}17 & 1243.65 & \hspace{0.5em}2148.53 & \hspace{0.5em}3.80 & 0.10 & 35.76 & \hspace{1em}37.79 \\
    informational & \hspace{0.9em}4 & 1561.75 & \hspace{0.5em}4124.25 & \hspace{0.5em}6.41 & 0.11 & \hspace{0.5em}8.00 & \hspace{0.5em}174.06\\
    \end{tabularx}
    \caption{Summary statistics of the network corpus used to evaluate algorithms for the Information Access Gap Minimization problem, showing the number of networks by scientific domain, along with the average number of nodes $\langle n \rangle$, average number of edges $\langle m \rangle$, average degree $\langle k \rangle$, average clustering coefficient (transitivity) $\langle C \rangle$, average diameter $\langle \ell_{\rm max} \rangle$, and average variance of the degree distribution $\langle \sigma^2 \rangle$ for networks in that domain.}
    \label{table:corpus_stats}
\end{table*}
\section{Network Corpus}
Summary statistics for the network corpus can be found in Table~\ref{table:corpus_stats}. 

\section{Pseudocode for Algorithms}
For completeness, here we provide pseudocode for all 10 new algorithms, along with high-level descriptions. We also describe the ensemble and meta-learner approaches.\\

\noindent \textbf{Algorithm 1:\ Myopic BFS}\\
Approximates $\pi_i$ values by performing BFS traversal from the most recent seed and computing $\pi_i$ activation probabilities for every node in the network, taking into account only its ``parents'' and ``neighbors.'' For a node $x$ at distance $t$ from the seed, a ``parent'' is any node at distance $t-1$ from the seed that shares an edge with $x$, and a ``neighbor'' is any node at distance $t$ that shares an edge with $x$. When adding a new seed, update the old probabilities with a new BFS traversal.

\medskip \noindent
    \begin{algorithm}
        \Input{$G = (V, E), s, \alpha$.}
        \Output{$S \subseteq V$ with $|S| = s+1$.}
        \caption{\label{alg:Myopic-BFS}\texttt{Myopic BFS}}
        \BlankLine
        \tcp{Initialize $S$ with a random node.}
        $S = \{s_0 \in_R V\}$ \\
        \While{$|S| \leq s$}{
            $Q \gets S[-1]$ \\ \tcp{initialize queue with latest seed}
            
            \While{$|Q| < 0$}{
                Perform BFS traversal and record distance $t$ to $S[-1]$ for every node
            }

            \For{$v \in V$}{
                Approximate access probability $\pi_i \forall v \in V$ \tcp{see the writeup earlier for more info}
            }

            \tcp{Choose the node with min. activation prob.}
            $x = \argmin_{v \in V \setminus S}\texttt{Prob}(v)$\\
            $S = S \cup \{x\}$
        }
        \Return $S$
    \end{algorithm}

\noindent
    \begin{algorithm}
        \Input{$G = (V, E), s$.}
        \Output{$S \subseteq V$ with $|S| = s+1$.}
        \caption{\label{alg:Naive-Myopic}\texttt{Naive Myopic BFS}}
        \BlankLine
        \tcp{Initialize $S$ with a random node.}
        $S = \{s_0 \in_R V\}$ \\
        
        $Q \gets S[-1]$ \\ \tcp{initialize queue with first random seed}
        
        \While{$|Q| < 0$}{
            Perform BFS traversal and record distance $t$ to $S[-1]$ for every node
        }

        \For{$v \in V$}{
            Approximate access probability $\pi_i \forall v \in V$ \tcp{see the writeup earlier for more info}
        }

        \tcp{Choose $s$ nodes with min. activation prob.}
        \While{$|S| \leq s$}{
            $x = \argmin_{v \in V \setminus S} \texttt{Prob}(v)$ \\
            $S = S \cup \{x\}$ \\
        }
    \Return $S$
    \end{algorithm}
\noindent \textbf{Algorithm 2: Naive Myopic BFS}\\ 
Approximates $\pi_i$ values in the same way as \texttt{Myopic BFS}. Reuses the approximations from the first traversal when adding each subsequent seed.

\noindent \textbf{Algorithm 3:\ Myopic PPR}\\
Choose an initial seed node uniformly at random. Then, for each new seed, perform a pass of Personalized Page Rank, with personalization parameter set to bias random walk restarts from the current seed set. Sort the nodes by PPR in ascending order, choose $s$ lowest-scoring nodes as new seeds.

\medskip \noindent
    \begin{algorithm}
        \Input{$G = (V, E), s$.}
        \Output{$S \subseteq V$ with $|S| = s+1$.}
        \caption{\label{alg:Myopic-PPR}\texttt{Myopic PPR}}
        \BlankLine
        \tcp{Initialize $S$ with a random node.}
        $S = \{s_0 \in_R V\}$ \\
        \While{$|S| \leq s$}{
            $x = \argmin_{v \in V \setminus S} \texttt{PPR}(v, S)$ \\
            $S = S \cup \{x\}$ \\
        }
    \Return $S$
    \end{algorithm}
\bigskip

\noindent \textbf{Algorithm 4:\ Naive Myopic PPR}\\ 
Choose an initial seed node uniformly at random.  Then, perform a pass of Personalized Page Rank, with personalization parameter set to bias random walk restarts from the initial random seed. Sort the nodes by PPR in ascending order, choose $s$ lowest-scoring nodes as new seeds.

\medskip \noindent
    \begin{algorithm}
        \Input{$G = (V, E), s$.}
        \Output{$S \subseteq V$ with $|S| = s+1$.}
        \caption{\label{alg:Naive-Myopic-PPR}\texttt{Naive Myopic PPR}}
        \BlankLine
        \tcp{Initialize $S$ with a random node.}
        $S = \{s_0 \in_R V\}$ \\
        \tcp{Sort the nodes by ascending \texttt{PPR}}
        $v_1, v_2, \ldots, v_n = \texttt{PPR}(G, S)$\\
        $S = S \cup \{v_1, v_2, \ldots v_s\}$ \\
    \Return $S$
    \end{algorithm}
\bigskip
\bigskip

\noindent \textbf{Algorithm 5:\ LeastCentral}\\
Choose an initial seed node uniformly at random. Then choose the node with the lowest closeness centrality as the new seed.

\noindent
    \begin{algorithm}
        \Input{$G = (V, E), s$.}
        \Output{$S \subseteq V$ with $|S| = s+1$.}
        \caption{\label{alg:LeastCentral}\texttt{LeastCentral}}
        \BlankLine
        \tcp{Initialize $S$ with a random node.}
        $S = \{s_0 \in_R V\}$ \\
        \While{$|S| \leq s$}{
        \tcp{Choose node with min. Close. Centrality}
        $x = \argmin_{v \in V \setminus S} \texttt{CC}(v)$ \\
        $S = S \cup \{x\}$\\
        }
    \Return $S$
    \end{algorithm}

\noindent \textbf{Algorithm 6:\ LeastCentral\_n}\\ 
Choose an initial seed node uniformly at random. Then choose a node with the smallest closeness centrality and select that node's highest-degree neighbor to be the next seed.

\medskip \noindent
    \begin{algorithm}
        \Input{$G = (V, E), s$.}
        \Output{$S \subseteq V$ with $|S| = s+1$.}
        \caption{\label{alg:LeastCentral-n}\texttt{LeastCentral\_n}}
        \BlankLine
        \tcp{Initialize $S$ with a random node.}
        $S = \{s_0 \in_R V\}$ \\
        \While{$|S| \leq s$}{
            \tcp{Choose node with min. Close. Centrality}
            $x = \argmin_{v \in V \setminus S} \texttt{CC}(v)$ \\
            \tcp{Choose the highest degree neighbor.}
            $y = \argmax_{v \in N(x)} d(v)$ \\
            $S = S \cup \{y\}$\\
        }
    \Return $S$
    \end{algorithm}
\bigskip

\noindent \textbf{Algorithm 7:\ MinDegree\_hc}\\ 
Choose an initial seed node uniformly at random. Then, identify all minimum degree nodes, and sort by harmonic centrality. Finally, choose the node with the lowest harmonic centrality as the new seed.

\medskip \noindent
    \begin{algorithm}
        \Input{$G = (V, E), s$.}
        \Output{$S \subseteq V$ with $|S| = s+1$.}
        \caption{\label{alg:MinDegree-hc}\texttt{MinDegree\_hc}}
        \BlankLine
        \tcp{Initialize $S$ with a random node.}
        $S = \{s_0 \in_R V\}$ \\
        \While{$|S| \leq s$}{
            \tcp{Choose nodes with min. degree.}
            $V' = \{v \in V\setminus S \colon d(w) \geq d(v) \ \forall w \in V \setminus S\}$ \\
            \tcp{Choose node with min. Harm. Centrality.}
            $x = \argmin_{v \in V'} \texttt{HC}(v)$ \\
            $S = S \cup \{x\}$ \\ 
        }
    \Return $S$
    \end{algorithm}
\bigskip

\noindent \textbf{Algorithm 8:\ MinDegree\_hcn}\\ 
Choose an initial seed node uniformly at random. Then, identify all minimum degree nodes, and sort by harmonic centrality. Finally, choose the highest-degree neighbor of the node with the lowest harmonic centrality as the new seed.

\medskip \noindent
    \begin{algorithm}
        \Input{$G = (V, E), s$.}
        \Output{$S \subseteq V$ with $|S| = s+1$.}
        \caption{\label{alg:MinDegree-hcn}\texttt{MinDegree\_hcn}}
            \BlankLine
            \tcp{Initialize $S$ with a random node.}
          $S = \{s_0 \in_R V\}$ \\
          \While{$|S| \leq s$}{
            \tcp{Choose nodes with min. degree.}
            $V' = \{v \in V\setminus S \colon d(w) \geq d(v) \ \forall w \in V \setminus S\}$ \\
            \tcp{Choose node with min. Harm. Centrality.}
            $x = \argmin_{v \in V'} \texttt{HC}(v)$ \\
            \tcp{Choose highest degree neighbor.}
            $y = \argmax_{v \in N(x)} d(v)$ \\
            $S = S \cup \{y\}$ \\
        \Return $S$ }
    \end{algorithm}
\bigskip
\vfill\eject
\noindent \textbf{Algorithm 9:\ MinDegree\_nd}\\
Choose an initial seed node uniformly at random. Then, identify all minimum degree nodes, and sort by neighbor degree. Choose the node with the highest-degree neighbor as the new seed.

\medskip \noindent
    \begin{algorithm}
        \Input{$G = (V, E), s$.}
        \Output{$S \subseteq V$ with $|S| = s+1$.}
        \caption{\label{alg:MinDegree-nd}\texttt{MinDegree\_nd}}
        \BlankLine
        \tcp{Initialize $S$ with a random node.}
        $S = \{s_0 \in_R V\}$ \\
        \While{$|S| \leq s$}{
            \tcp{Choose nodes with min. degree.}
            $V' = \{v \in V\setminus S \colon d(w) \geq d(v) \ \forall w \in V \setminus S\}$ \\
            \tcp{Choose node with highest neigh. degree.}
            $x = \argmax_{v \in V'} \sum_{w \in N(v)} d(w)$ \\
            $S = S \cup \{x\}$\\

        }
    \Return $S$
    \end{algorithm}
\bigskip

\noindent \textbf{Algorithm 10:\ MinDegree\_ndn}\\
Choose an initial seed node uniformly at random. Then, identify all minimum degree nodes, and sort by neighbor degree. For the node with the highest-degree neighbor, choose the highest-degree neighbor itself as the new seed.

\medskip \noindent
    \begin{algorithm}
        \Input{$G = (V, E), s$.}
        \Output{$S \subseteq V$ with $|S| = s+1$.}
        \caption{\label{alg:MinDegree-ndn}\texttt{MinDegree\_ndn}}
        \BlankLine
        \tcp{Initialize $S$ with a random node.}
        $S = \{s_0 \in_R V\}$ \\
        \While{$|S| \leq s$}{
            \tcp{Choose nodes with min. degree.}
            $V' = \{v \in V\setminus S \colon d(w) \geq d(v) \ \forall w \in V \setminus S\}$ \\
            \tcp{Choose node with highest neigh. degree.}
            $x = \argmax_{v \in V'} \sum_{w \in N(v)} d(w)$ \\
            \tcp{Choose highest degree neighbor.}
            $y = \argmax_{v \in N(x)} d(v)$ \\
            $S = S \cup \{y\}$ \\
        }
    \Return $S$
    \end{algorithm}
\bigskip
\vfill\eject

\noindent \textbf{Algorithm 11:\ Fast Ensemble (Oracle)}\\
Select five members of $D$, an ensemble of algorithms. To do so, perform greedy search for a set of five algorithms that attain $\geq 80\%$ of \texttt{Myopic}'s performance on the network corpus, as evaluated by \texttt{ProbEst}. Then, for each new network, select the algorithm that performs best as suggested by the oracle, and apply it.

\medskip \noindent
    \begin{algorithm}
        \Input{$G = (V, E), s$.}
        \Output{$S \subseteq V$ with $|S| = s+1$.}
        \caption{\label{alg:fast-oracle}\texttt{Fast Ensemble (Oracle)}}

        Initialize $S$ with a random node.\;
        $S = \{s_0 \in_R V\}$
        
        Select $a_1,a_2,a_3,a_4,a_5$ to maximize number of networks in the corpus for which performance of some $a_i$ is $\geq 80\%$ of Myopic\;
        $D \gets \{a_1,a_2,a_3,a_4,a_5\}$
        
        \tcp{Ask the oracle $O$ which algorithm to use}
        $a_{\rm best} \gets O(G,D)$

        \tcp{Run the algorithm to get $S$}
        $S = S \cup a_{\rm best}(G,s)$

        \Return $S$        
    \end{algorithm}
\bigskip

\noindent \textbf{Algorithm 12:\ Meta-learner}\\
Select five members of $D$, an ensemble of algorithms. To do so, perform greedy search for a set of five algorithms that attain $\geq 80\%$ of \texttt{Myopic}'s performance on the network corpus, as evaluated by \texttt{ProbEst}. Using prior network corpus data, train a model $M$ that, given a new network $G$, selects the best algorithm out of the ensemble based on topological features. Apply the selected algorithm to $G$.

\medskip \noindent
    \begin{algorithm}
        \Input{$G = (V, E), s$.}
        \Output{$S \subseteq V$ with $|S| = s+1$.}
        \caption{\label{alg:metalearner}\texttt{Meta-learner}}

        Initialize $S$ with a random node.\;
        $S = \{s_0 \in_R V\}$
        
        Select $a_1,a_2,a_3,a_4,a_5$ to maximize number of networks in the corpus for which performance of some $a_i$ is $\geq 80\%$ of Myopic\;
        $D \gets \{a_1,a_2,a_3,a_4,a_5\}$

        Train Random Forest Classifier $M$ to select the best algorithm from $D$ based on network topology, using data from the network corpus\;
        
        \tcp{Ask $M$ which algorithm to use}
        $a_{\rm best} \gets M(G,D)$

        \tcp{Run the algorithm to get $S$}
        $S = S \cup a_{\rm best}(G,s)$

        \Return $S$  
    \end{algorithm}

\begin{figure*}[t!]
    \begin{subfigure}{0.9\textwidth}
    \includegraphics[width=1\textwidth]{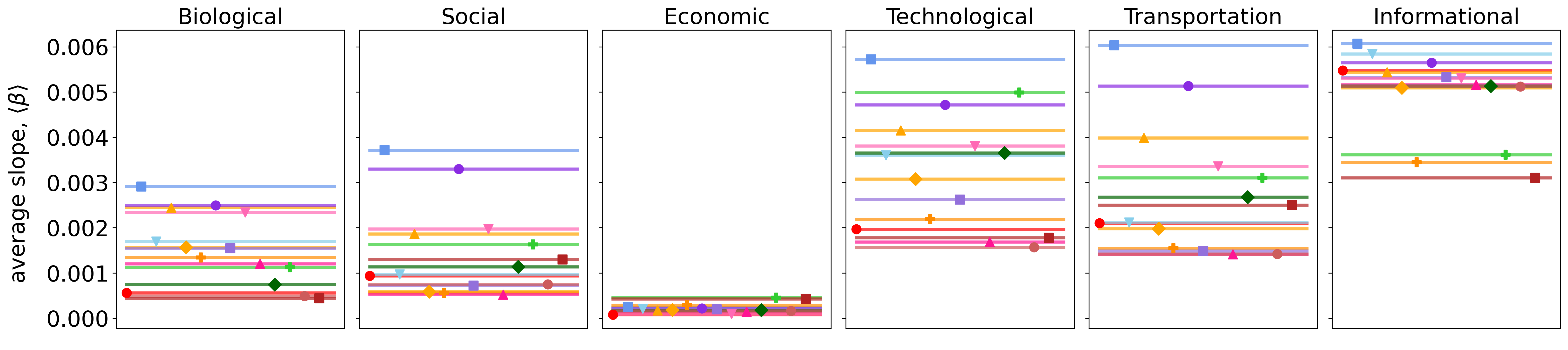}
    \vspace*{-10mm}
    \caption{}
    \label{fig:algo_mean_performance_low}
    \end{subfigure}

    \vspace*{0mm}

    \begin{subfigure}{0.9\textwidth}
    \includegraphics[width=1\textwidth]{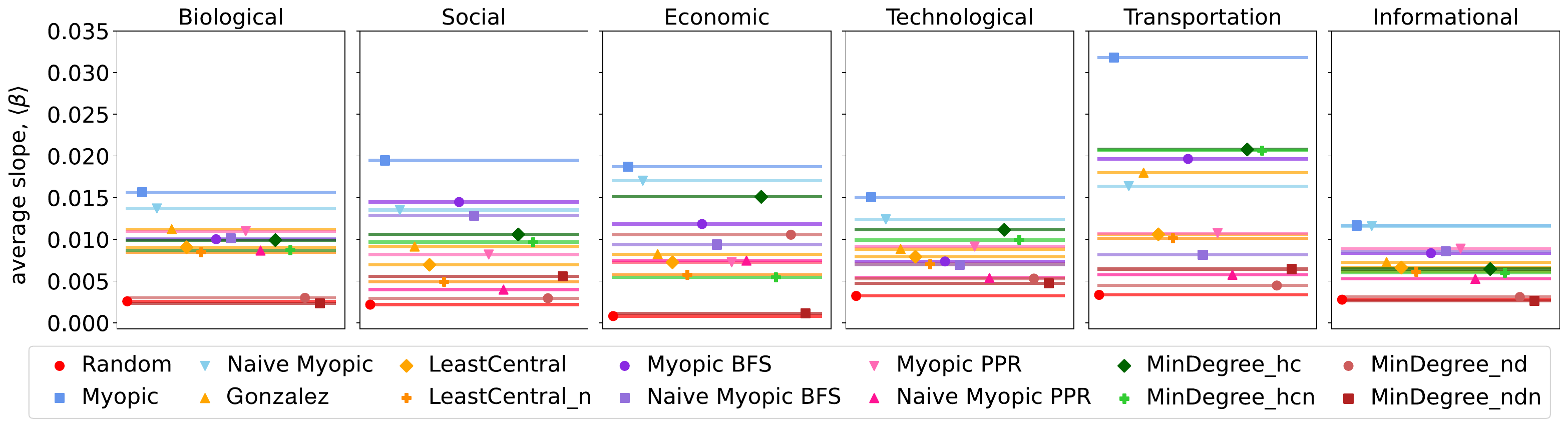}
    \caption{}
    \end{subfigure}
    
    \caption{Mean performance of the intervention algorithms on each domain in the corpus, under (a) low spreadability and (b) high spreadability settings. Medium spreadability results are given in~\Cref{fig:algo_mean_performance}.}
    \label{fig:algo_mean_performance_high}
\end{figure*}

\begin{figure*}[t!]
    \begin{subfigure}{0.9\textwidth}
    \includegraphics[width=1\textwidth]{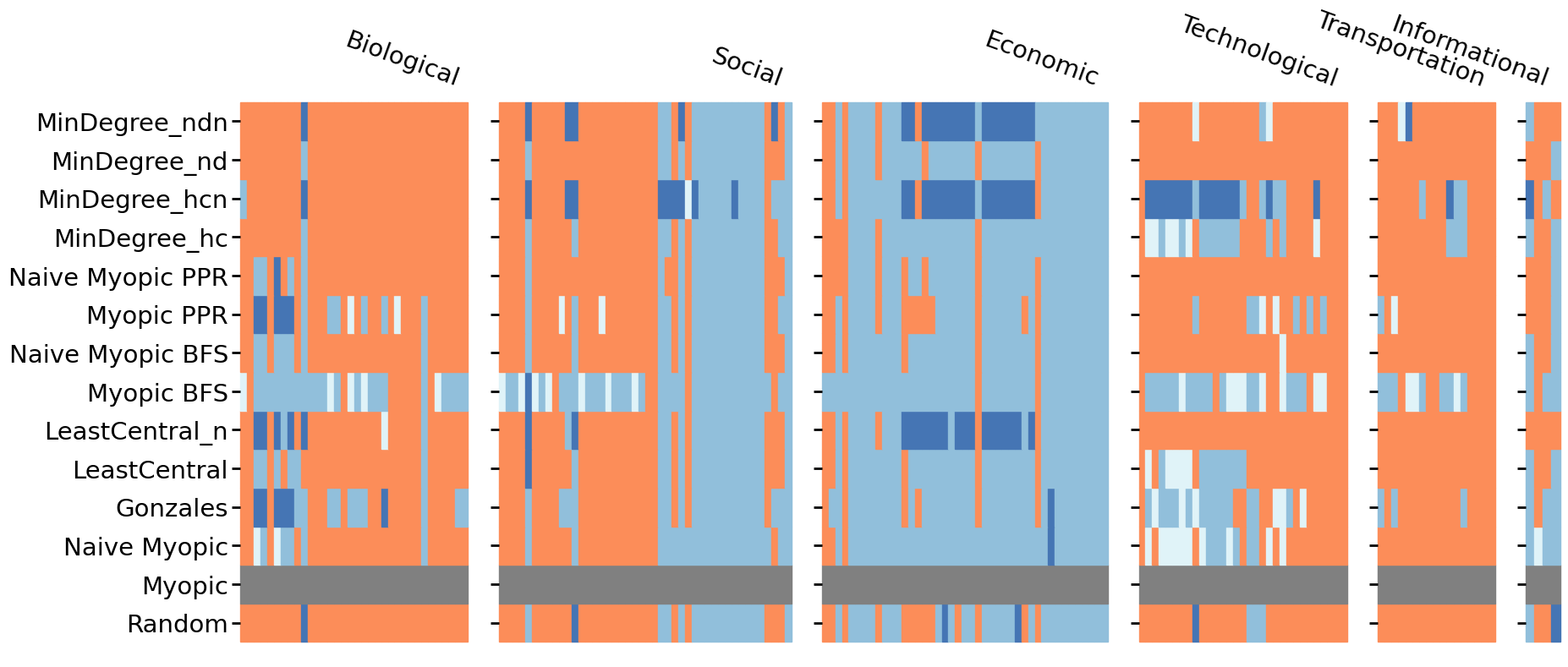}
    \vspace*{-10mm}
    \caption{}
    \label{fig:algo_performance_heatmap_low}
    \end{subfigure}

    \vspace*{2mm}

    \begin{subfigure}{0.9\textwidth}
    \includegraphics[width=1\textwidth]{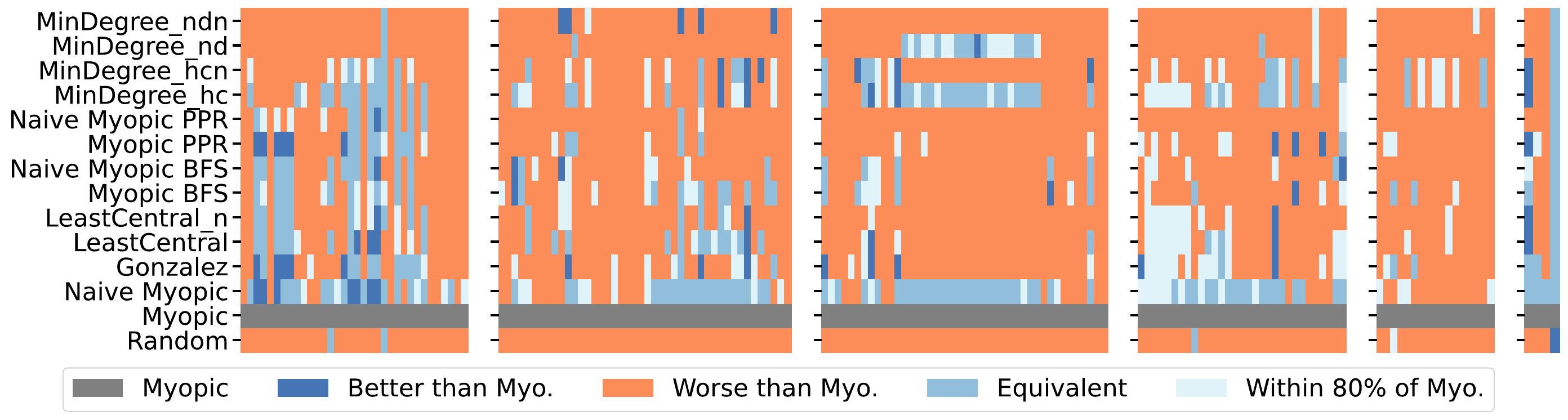}
    \vspace*{-19mm}
    \caption{}
    \label{fig:algo_performance_heatmap_high}
    \end{subfigure}

    \vspace*{8mm}

    \caption{Performance of intervention algorithms on the network corpus, relative to \texttt{Myopic} and sorted in ascending order by network size within each domain. (a): low spreadability, $12\%$ of networks have no algorithm better than or within $80\%$ of \texttt{Myopic}'s performance; (b): high spreadability, showing very similar results to medium spreadability (\Cref{fig:algo_performance_heatmaps}). ``Equivalent'' defined as within one standard error of $\beta$ for \texttt{Myopic}; typically about 0.001.}
    \label{fig:supp_algo_performance_heatmaps}
\end{figure*}
\section{Supplementary Results}\label{appendix:supplementary-results}
\paragraph{Algorithm performance for low and high spreadability.}
Figure~\ref{fig:algo_mean_performance_high} shows the mean performance $\beta$ of all the algorithms applied to all the networks in the corpus, grouped into the six network domains:\ biological, social, economic, technological, transportation, and informational, for low and high spreadability settings. We observe qualitatively similar results across spreadability settings, in which \texttt{Myopic} is often on average the best performing algorithm. However, the relative ordering of other algorithms varies substantially across domains and spreadability settings.

Figure~\ref{fig:supp_algo_performance_heatmaps} expands on these results (following the experimental steps in the main text), showing for the low and high spreadability settings the performance of each algorithm relative to the performance of \texttt{Myopic}. Notably, in the low spreadability setting, many more algorithms fall into the ``equivalent'' category, in which their performance is statistically indistinguisable from \texttt{Myopic} ($\beta$ within one standard error). This behavior is due to the low precision of \texttt{ProbEst} in this setting. We find many fewer cases of equivalence in the high spreadability setting, which is qualitatively similar to the results in the medium setting, in which across networks, many alternatives perform similarly as \texttt{Myopic}, and in some cases outperform \texttt{Myopic}. These results reinforce the finding that generally no algorithm is superior to others across networks and spreadability settings.

\begin{figure}[t!]
    \includegraphics[width=0.45\textwidth]{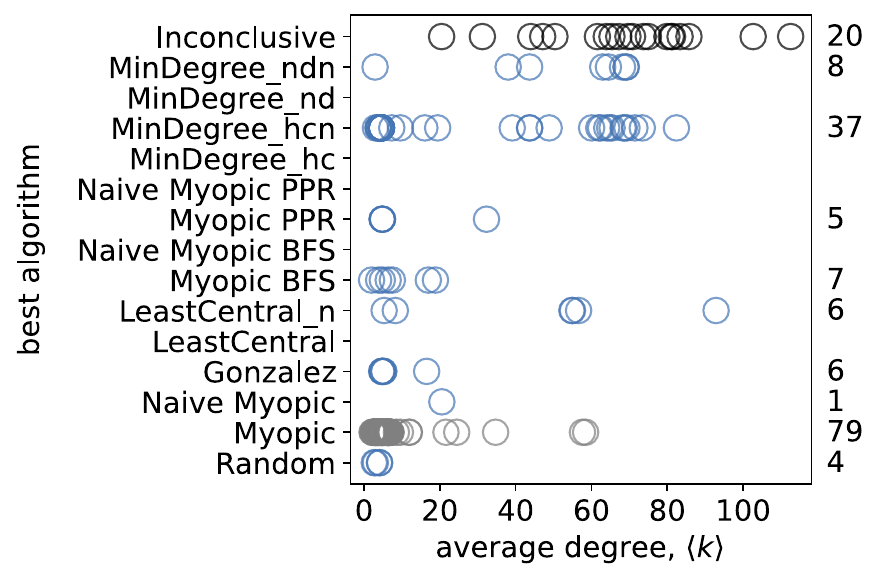}
    \caption{Best-performing algorithm on a network vs average degree of the network under low spreadability. Each network's corresponding single most-performant algorithm is plotted with a circle. Total number of circles per line is given on the right-hand side, indicating how many times an algorithm was the best-performing one across the entire corpus.}
    \label{fig:best_vs_avg_degree_low}
\end{figure}

\begin{figure}[t!]
    \includegraphics[width=0.45\textwidth]{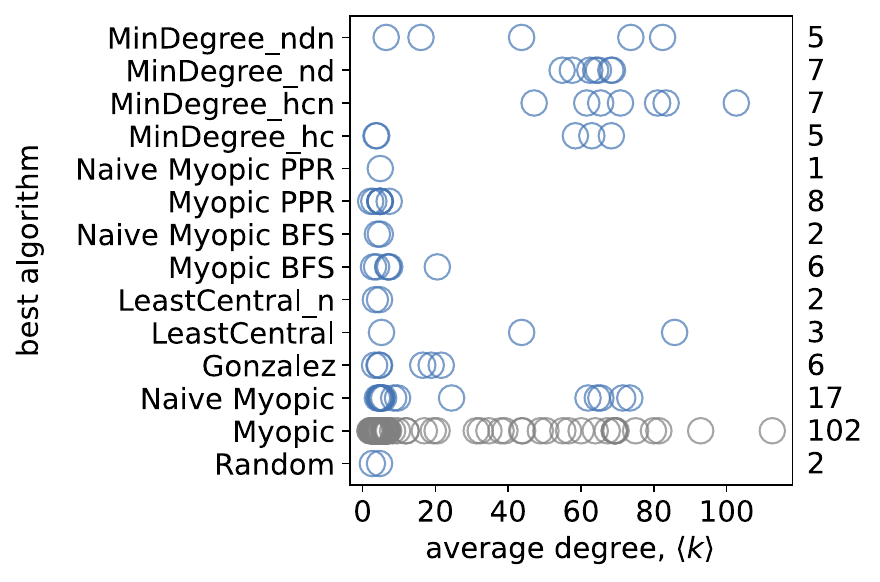}
    \caption{Best-performing algorithm on a network vs average degree of the network under high spreadability. Each network's corresponding single most-performant algorithm is plotted with a circle. Total number of circles per line is given on the right-hand side, indicating how many times an algorithm was the best-performing one across the entire corpus.}
    \label{fig:best_vs_avg_degree_high}
\end{figure}

Figures~\ref{fig:best_vs_avg_degree_low} and~\ref{fig:best_vs_avg_degree_high} tabulate the counts of how often a particular algorithm was the best performing (highest $\beta$) and plot them as a function of network mean degree $\langle k \rangle$, for low and high spreadability, respectively. Here we see clearly that \texttt{Myopic} is by far the best algorithm in high spreadability (similar to results for medium, shown in~\Cref{fig:best_vs_avg_degree}), but is much less so under low spreadability. We believe this difference is attributable to the imprecision of \texttt{ProbEst} in the low spreadability setting.

\begin{figure}[t!]
\centering
\includegraphics[width=0.35\textwidth]{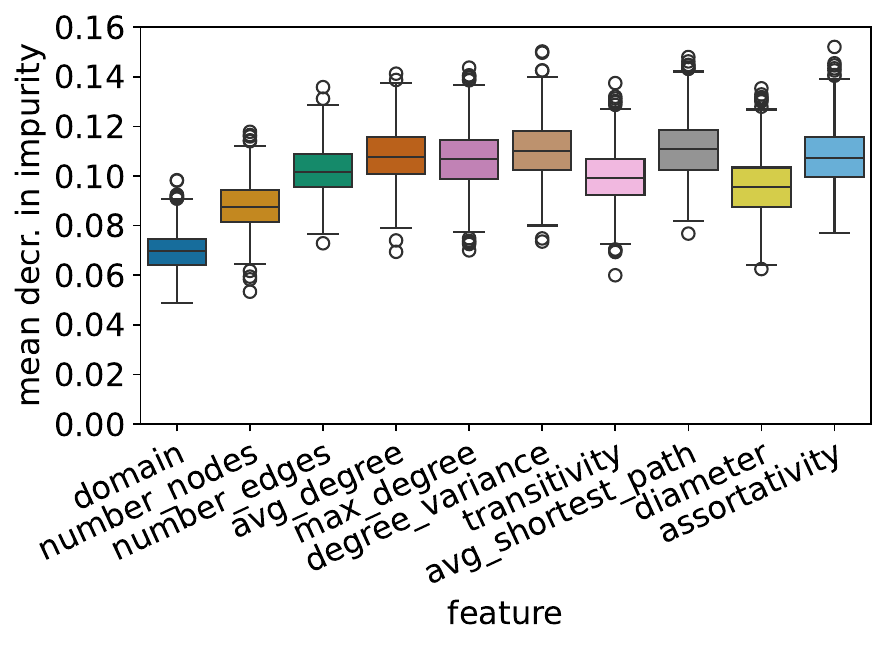}
\caption{Meta-learner random forest feature importance for predicting which of five algorithms is the best choice for a given network. Higher values mean a feature contributes more to the predicted performance value.}
\label{fig:ml_importances}
\end{figure}

\begin{figure}[t!]
    \includegraphics[width=0.45\textwidth]{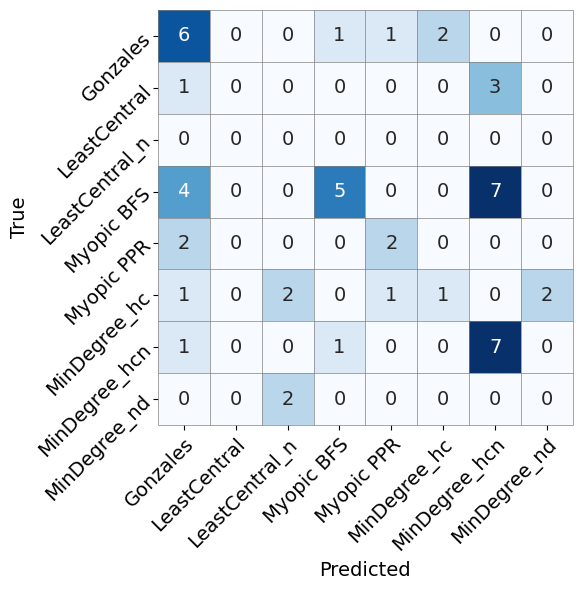}
    \caption{Confusion matrix of the best algorithm prediction model for the meta-learner.}
    \label{fig:ml_confusion}
\end{figure}
\paragraph{Meta-learner.}
To construct the meta-learner, we trained a model to predict which algorithm would perform best on a given network, given only the network's features and the spreadability of the information in the Independent Cascade model. For this task, we held out \texttt{Myopic} and \texttt{Naive Myopic}, as the goal is to approximate their state-of-the-art performance using more scalable algorithms. Network features used are:\ the network domain, number of nodes $n$, average degree $\langle k \rangle$, maximum degree $k_{\max}$, degree variance $\sigma^2$, clustering coefficient $C$, average shortest path $\langle \ell \rangle$, diameter $\ell_{\max}$, and degree assortativity $r$. We use this approach rather than a representation learning approach to ensure better interpretability of what aspects of network structure influence performance predictability.

This multi-class classification task is challenging. Fine-tuned models come out with an accuracy of $\approx 0.51$. Figs.~\ref{fig:ml_importances} and~\ref{fig:ml_confusion} show the learned feature importances and confusion matrix for the medium spreadability setting (similar results for low and high spreadability models). However, generally, incorrect algorithm choices by the model tend to select an alternative that is almost as good as the true best, in terms of performance $\beta$, and so classification errors often do not translate into large losses in performance.

From the classification task, we find that the domain of a network contributes the least to the prediction accuracy, followed by number of nodes $n$. Other network features are roughly equally important (\Cref{fig:ml_importances}). This result implies that network domain does not contribute much marginal information beyond a network's features, as has been found previously by Ikehara and Clauset~\cite{ikehara2017}. That is, network features encode much the same information as network domain.

\begin{figure}[t!]
\includegraphics[width=0.45\textwidth]{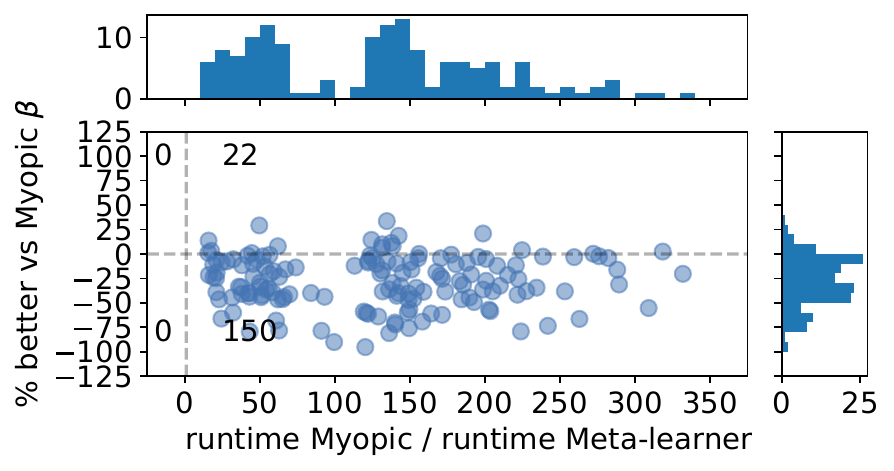}
    \caption{Performance difference vs.\ speedup for the meta-learning algorithm relative to \texttt{Myopic} under high spreadability, with marginal histograms. Results averaged over $1000$ meta-learner runs, and extreme outliers have been removed for visualization purposes only. Average performance loss relative to \texttt{Myopic} is $28.20 \pm 25.80$, with an average speedup factor of $133.35 \pm 79.32$, but for 22 networks (12.9\%), the meta-learner outperforms \texttt{Myopic}.
    }
\label{fig:metalearner_high}
\end{figure}

The accuracy of the meta-learning model under medium and high spreadability (Figs.~\ref{fig:metalearner_med} and~\ref{fig:metalearner_high}) are similar, but with slightly larger performance loss and substantially larger runtime speedup in the case of high, than in medium.

Unsurprisingly, the fast ensemble algorithms (which use an oracle to choose the best heuristic for a given network) in both medium and high spreadability settings outperform the meta-learner. However, the gap between fast ensemble and meta-learning is only about 10-20\%, indicating that while there is room for improvement in the meta-learner's classifier, the optimal performance is not far off. Moreover, the fast ensemble does outperform \texttt{Myopic} on more cases than does the meta-learning, but again, the gap is not enormous. We note that timing results for some algorithms (both those using \texttt{ProbEst} and the BFS-based class of new algorithms; see~\Cref{sec:algos}) may be impacted by our use of an adjacency matrix (rather than adjacency list) representation for all networks.

\begin{figure}[t!]
\includegraphics[width=0.45\textwidth]{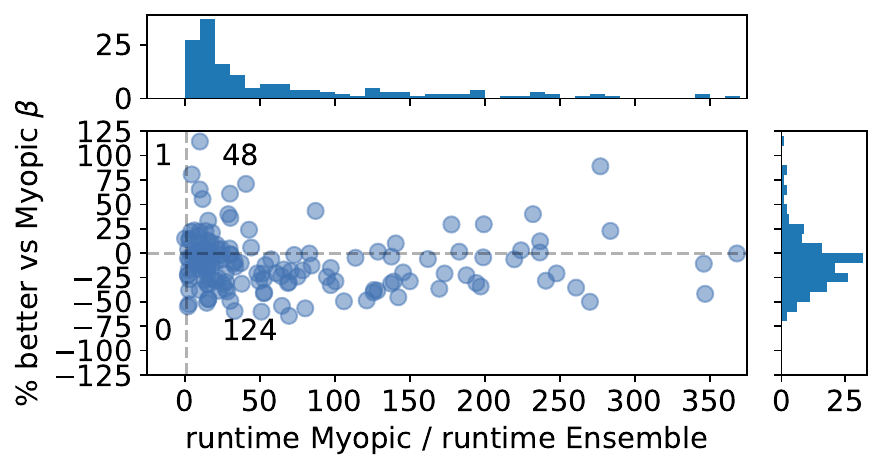}
    \caption{Performance difference vs.\ speedup for the fast ensemble algorithm relative to \texttt{Myopic} under medium spreadability, with marginal histograms. Extreme outliers have been removed for visualization purposes only. Average performance loss relative to \texttt{Myopic} is $9.34 \pm 28.34\%$, with an average speedup factor of  $70.66 \pm 85.65$, but for 48 networks (28.1\%), the fast ensemble outperforms \texttt{Myopic}.}
\label{fig:ensemble_med}
\end{figure}

\begin{figure}[t!]
\includegraphics[width=0.45\textwidth]{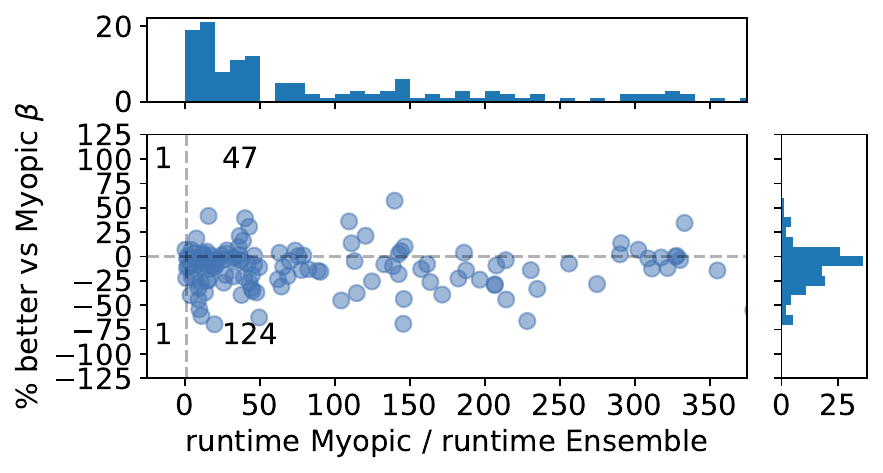}
    \caption{Performance difference vs.\ speedup for the fast ensemble algorithm relative to \texttt{Myopic} under high spreadability, with marginal histograms. Extreme outliers have been removed for visualization purposes only. Average performance loss relative to \texttt{Myopic} is $11.11 \pm 21.8\%$, with an average speedup of $101.88 \pm 101.27$, but for 47 networks (27.6\%), the fast ensemble algorithm outperforms \texttt{Myopic}.}
\label{fig:ensemble_high}
\end{figure}

\pagebreak 

\section{Results on Larger Networks}
\label{appendix:large_networks}

\begin{table}[h]
\begin{tabular}{l|l|l|l|l|l|l}
               & \multicolumn{1}{c|}{$n$} & \multicolumn{1}{c|}{$m$} & \multicolumn{1}{c|}{$k$} & \multicolumn{1}{c|}{$C$} & $l_{max}$ & $\sigma^2$ \\ \hline
Email (EU)  & 34,203                 & 96,669                 & 15.87                  & 0.04  &   7     & 770   \\ \hline
Google+ (2013)        & 86,785                 & 688,602                 & 5.65                   & 0.27  &   47  & 1,809      \\ \hline
\end{tabular}
\caption{\label{table:large-networks} Statistics of Email (EU) and Google+ (2013) showing number of nodes $n$, number of edges $m$, average node degree $k$, clustering coefficient $C$, diameter $l_{max}$, and the variance $\sigma^2$ of the degree distribution.}
\end{table}

\begin{figure}[t!]
    \centering
    \includegraphics[width=1\linewidth]{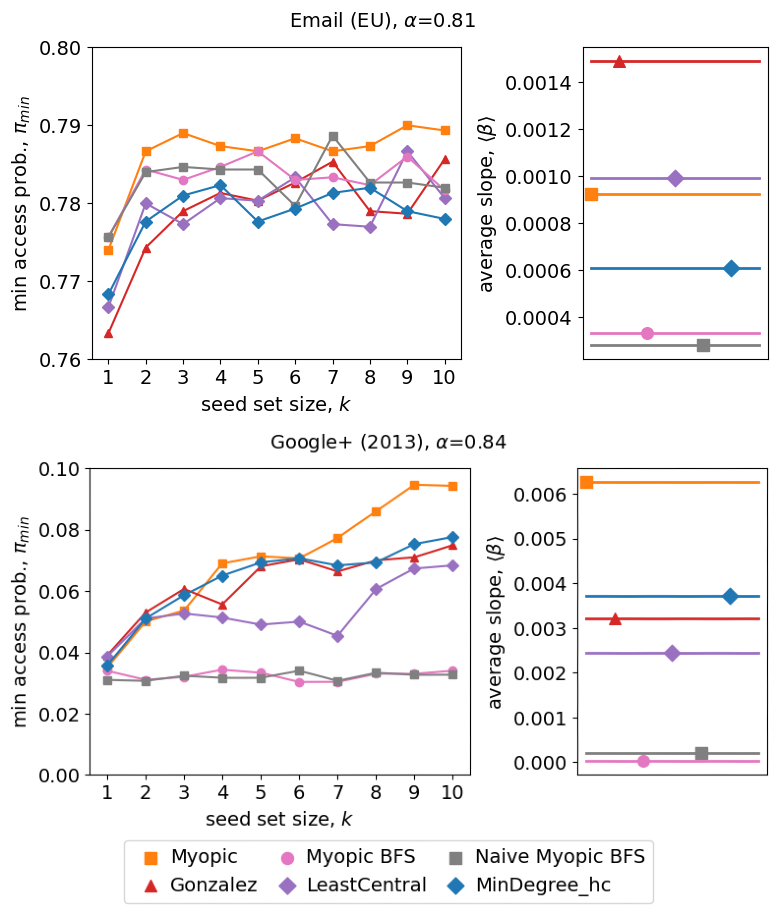}
    \caption{Minimum access probability vs. seed set size under the six algorithms in $\mathcal{X}_{\rm high}$, evaluated on the two larger networks, Email (EU) and Google+ (2013) under high spreadability. Small decreases in $\pi_{\text{min}}$ are discussed in~\Cref{appendix:large_networks}. The panel on the right shows $\beta$, the slope of the line of best fit, for each algorithm.}
    \label{fig:larger_networks_performance}
\end{figure}

In~\Cref{table:large-networks}, we present summary statistics for the two larger networks used in~\Cref{sec:large-networks}.
In~\Cref{fig:larger_networks_performance}, we evaluate the performance of the five algorithms from $\mathcal{X}_\text{high}$ as well as that of \texttt{Myopic}.
When evaluating the performance of these algorithms, it is important to note that due to the size of the networks, a large increase in $\pi_{\text{min}}$ is not expected with any algorithm when choosing only 10 seeds. Additionally, when looking at the minimum access probability as seeds are added, there are small decreases in $\pi_{\text{min}}$ resulting from the imprecise nature of $\texttt{ProbEst}$. This is a limitation on our ability to evaluate these algorithms, but it is not a restriction on the algorithms themselves. That being noted, three out of the five algorithms ($\texttt{MinDegree\_hc, Gonzalez}$ and  $\texttt{LeastCentral}$) produce $\beta$ values close to, and in some cases even greater than, $\texttt{Myopic}$ (\Cref{fig:larger_networks_performance}). Although the $\beta$ values are predictably low for all algorithms, including $\texttt{Myopic}$, there is a general positive trend with most algorithms. From these results, we hypothesize that when scaled to large networks, the speedup of the algorithms is sustained, while seed choice quality remains comparable to $\texttt{Myopic}$ for at least some algorithms.

\end{document}